\documentclass[11pt,a4paper]{article}

\usepackage{jcappub}
\usepackage[latin1]{inputenc}
\usepackage{bbm}
\usepackage{graphicx}
\usepackage{epsfig}
\usepackage{subfigure}
\usepackage{color}

\newcommand{\deldag}{\mathbin{\partial\mkern-10.5mu\big/}}
\newcommand{\Ddag}{\mathbin{D\mkern-11.5mu\big/}}

\newcommand{\be}{\begin{equation}}          
\newcommand{\ee}{\end{equation}}
\newcommand{\bea}{\begin{eqnarray}}         
\newcommand{\eea}{\end{eqnarray}}
\newcommand{\nn}{\nonumber}
\newcommand{\bmp}{\noindent\begin{minipage}{16cm}}
\newcommand{\emp}{\end{minipage}\vskip 7mm} 
\def\lsim{\mathrel{\raise.3ex\hbox{$<$\kern-.75em\lower1ex\hbox{$\sim$}}}}
\def\gsim{\mathrel{\raise.3ex\hbox{$>$\kern-.75em\lower1ex\hbox{$\sim$}}}}
\newcommand{\ie}{{\em i.e.~}}

\newcommand{\dsl}{\mathbin{\partial\mkern-10mu\big/}}
\newcommand{\intron}[1]{}

\def\sfrac#1#2{{\textstyle\frac{#1}{#2}}}
\def\ww{{\rm \scriptscriptstyle W}}

\def\MNN{M_{\scriptscriptstyle \rm NN}}
\def\Mww{M_{\scriptscriptstyle ww}}
\def\Mbb{M_{\scriptscriptstyle \beta \beta}}
\def\mNb{m_{\scriptscriptstyle \rm N \beta}}
\def\mwb{m_{\scriptscriptstyle w \beta}}
\def\mNw{m_{\scriptscriptstyle {\rm N}w}}
\def\LNN{\lambda_{\scriptscriptstyle \rm NN}}
\def\Lww{\lambda_{\scriptscriptstyle ww}}
\def\Lwb{\lambda_{\scriptscriptstyle w\beta}}

\def\mDM{m_{\scriptscriptstyle \rm DM}}
\def\ODM{\Omega_{\scriptscriptstyle \rm DM}}

\title{A model for dark matter, naturalness and a complete gauge unification}

\author[a,c]{Kimmo Kainulainen} 
\author[b,c]{Kimmo Tuominen} 
\author[a,c]{Jussi Virkaj\"arvi}

\affiliation[a]{Department of Physics, University of Jyv\"askyl\"a, \\
                      P.O.Box 35 (YFL), FI-40014 University of Jyv\"askyl\"a, Finland}
\affiliation[b]{Department of Physics, University of Helsinki, \\
                      P.O.Box 64, FI-00014 University of Helsinki, Finland}
\affiliation[c]{Helsinki Institute of Physics, \\
                     P.O.~Box 64, FI-00014 University of Helsinki, Finland}

\emailAdd{kimmo.kainulainen@jyu.fi}
\emailAdd{kimmo.i.tuominen@helsinki.fi}
\emailAdd{jussi.virkajarvi@jyu.fi}

\abstract{We consider dark matter in a minimal extension of the Standard Model (SM) which breaks electroweak symmetry dynamically and leads to a complete unification of the SM and technicolor coupling constants. The unification scale is determined to be $M_{\rm U} \approx 2.2 \times 10^{15}$ GeV and the unified coupling $\alpha_{\rm U} \approx 0.0304$. Moreover, unification strongly suggest that the technicolor sector of the model {\em must} become strong at the scale of ${\cal O}$(TeV). The model also contains a tightly constrained sector of mixing neutral fields stabilized by a discrete symmetry. We find the lightest of these states can be DM with a mass in the range $m_{\rm DM} \approx 30-800$ GeV. We find a large set of parameters that satisfy all available constraints from colliders and from dark matter search experiments. However, most of the available parameter space is within the reach of the next generation of DM search experiments. The model is also sensitive to a modest improvement in the measurement of the precision electroweak parameters.}

\keywords{Dark matter, Naturality, Unification, Technicolor \\
\\
\emph{Preprint: HIP-2015-12/TH}}

\begin{document}
\maketitle
%

%
\section{Introduction}
\label{Intro}
%

In contrast with its celebrated finding of the Higgs boson~\cite{Aad:2012tfa,Chatrchyan:2012ufa}, LHC has so far failed to discover any obvious sign for new physics. Yet compelling evidence for new physics abound. For example the existence of a large dark matter component in the energy density of the universe is currently lacking a proper elementary particle physics context. Standard model is also plagued by the hierarchy problem, \ie it lacks a natural explanation for the lightness of the Higgs particle. Finally, SM does not give rise to coupling constant unification, which would be highly desirable on theoretical grounds. Yet another issue is that Standard Model cannot explain why the universe contains only matter. Of these issues the DM problem is the most tangible one phenomenologically, and DM studies have indeed recently been the leading motivation for the construction and analyses of beyond SM scenarios. 

Dark matter can of course be considered separately from the issues of hierarchy problem and unification. In particular scalar dark matter models have been very popular recently~\cite{McDonald:1993ex,McDonald:2001vt,Burgess:2000yq,LopezHonorez:2006gr,Cline:2013gha,Cline:2013bln,Cline:2012hg,Alanne:2014bra}. These models are attractive because of their simplicity and because of possible connection to a new, larger dark sector interacting with the SM through the Higgs portal. Alternatively one can require that the DM is a part of a larger, independently motivated particle physics model. In this class the leading paradigm until recently was supersymmetry (SUSY) and in particular the minimal supersymmetric standard model. SUSY is compelling for many reasons, which include theoretical connections to (super)string theory and supergravity. From the phenomenological point of view SUSY is attractive because it could provide a solution to all main issues mentioned above. However, no sign of SUSY has been found. Instead the minimal supersymmetric standard model is getting ever more tightly constrained by the data. 

Another popular model building paradigm involves new strong dynamics (technicolor) sourcing the electroweak symmetry breaking. The currently favored technicolor (TC) models are based on the idea of quasiconformality \cite{Holdom:1984sk,Yamawaki:1985zg}, concrete realizations of which are the minimal and next to minimal walking technicolor models 
\cite{Sannino:2004qp,Dietrich:2005jn}.
Technicolor was originally put forward as a solution of the hierarchy problem. However, it was recently shown that TC models can easily also provide unification of SM gauge couplings~\cite{Kainulainen:2010pk,Kainulainen:2013sva,Gudnason:2006mk} and they have also shown to contain several possible DM candidates~\cite{Nussinov:1985xr,Chivukula:1989qb,Barr:1990ca,Gudnason:2006yj,Kainulainen:2006wq,Kouvaris:2007iq,Khlopov:2007ic,Khlopov:2008ty,Ryttov:2008xe,Foadi:2008qv,Kainulainen:2009rb,Kainulainen:2010pk,Kainulainen:2013sva,Hapola:2013mba}.

In this paper we consider a model for particle DM, which is deeply motivated by the TC paradigm and by the requirement of the gauge coupling unification. Yet, the DM-sector of the model may be considered also independent of its TC context. Therefore we first discuss the generic low-energy setup featuring new leptons transforming as a doublet and a triplet under the weak gauge group, and their mixing patterns and couplings with the gauge fields and with a Higgs-like scalar sector. The setup naturally leads to a thermal relic DM candidate, completely external to the strongly interaction sector and with fully deterministic couplings with ordinary matter. We also discuss a concrete model for dynamical electroweak symmetry breaking, which determines all mass terms and effective Higgs couplings. 

This model for dark matter was originally introduced in refs.~\cite{Kainulainen:2006wq,Kainulainen:2009rb,Kainulainen:2010pk,Kainulainen:2013sva}. Here we modify the underlying implementation of the model and significantly extend the analysis of ref.~\cite{Kainulainen:2013sva} by inclusion of all latest observational constraints. We also study more closely some benchmark scenarios for which we perform a new MCMC scan of the parameter space.
We also analyze new consistency constraints arising from LHC-bounds on invisible and radiative Higgs decays. We find that the model can provide a naturally stable DM particle with a mass in the range $m_{\rm DM} \sim 30 - 800$ GeV. Most of the allowed parameter space is within the reach of the next generation of direct and indirect DM search experiments. Also electroweak precision data, especially the Peskin-Takeuchi $S$-parameter, provide stringent bounds on the model.

We also show that the model gives rise to perfect 1-loop unification of {\em all} gauge couplings, including also the new technicolor interaction. The common unification scale and coupling are accurately determined: $M_{\rm U} \approx 2.2 \times 10^{15}$ GeV and $\alpha_{\rm U} \approx 0.0304$. The unification of the technicolor coupling is very sensitive to, and hence accurately determines the TC 1-loop IR-pole: $\Lambda_{\rm TC}^{\rm 1-loop} \approx 340$ GeV. Based on the QCD-analogue, this value strongly supports the natural TC scale $\Lambda_{\rm TC} \approx 3$ TeV. This result provides an important check on the internal consistency of the model.

The paper is organized as follows: We briefly review the model in section~\ref{sec:model} complete with determination of all gauge- and scalar couplings and the mass terms and a full implementation in the TC context. In section~\ref{sec:Unification} we study the unification of the standard model and the technicolor gauge couplings. The dark matter analysis together with imposing laboratory constraints is described in section~\ref{sec:Analysis}. Here we also show the results MCMC-scan of the model parameter space and show that our WIMP is compatible with all current observational constraints. In section~\ref{sec:Hgg} we consider the LHC constraints on higgs decays and radiative corrections to its mass. Finally in~\ref{sec:conclusions} we give our conclusions and outlook.

%
\section{General model of dark matter}
\label{sec:model}
%

We consider a model where the dark matter candidate is a mixture between electroweak singlet and neutral components of electroweak doublet and triplet fermions. Such candidate arises from a low energy Lagrangian of the form
\begin{equation}
{\mathcal{L}_{\rm{GM}}} = {\mathcal{L}_{4{\rm f,g}}} + {\mathcal{L}_{\rm{Ad},g}} +
{\mathcal{L}_{4{\rm f,H}}} + {\mathcal{L}_{\rm{Ad},H}} + {\mathcal{L}_{\rm{SM}}} \,,
\label{eq:Ltot}
\end{equation}
where ${\mathcal{L}_{4\rm{f,g}}}$ describes the gauge sector of heavy $4^{\rm{th}}$ lepton family and ${\mathcal{L}_{\rm{Ad,g}}}$ that of the SU(2) adjoint and singlet Weyl fermions and ${\mathcal{L}_{4\rm{f,H}}}$ and ${\mathcal{L}_{\rm{Ad,H}}}$ detail their interactions with scalar fields. Finally ${\mathcal{L}_{\rm{SM}}}$ describes Standard Model, where Higgs may be either a fundamental or an effective doublet field. We will now briefly introduce the relevant terms in the Lagrangian following ref.~\cite{Kainulainen:2013sva}.

%
\subsection{Gauge interactions}
%

We denote the left handed heavy doublet by $L_L = (N_L E_L)^T$ and the charged right handed singlet by $E_R$. Using the SM-like hypercharge assignments the kinetic and gauge interaction terms for these fields become equal to the corresponding terms in the SM:
\begin{eqnarray}
{\mathcal{L}_{4\rm{f,g}}}  
\,=\,  i\bar{L}_L \dsl L_L \,+\, i\bar{E}_R \dsl E_R \,
\, +\, {\mathcal{L}_{\rm{W}}} \,+\, {\mathcal{L}_{\rm{Z}}} \,+\, {\mathcal{L}_{\rm{A}}} \,,
\label{eq:Llep}
\end{eqnarray}
where the gauge currents are given by
\begin{eqnarray}
 {\mathcal{L}_{\rm{W}}} &=& 
   \frac{g}{\sqrt{2}} \left( W_{\mu}^{-} \bar{E}_L \gamma^{\mu} N_L + W_{\mu}^{+}\bar{N}_L 
                            \gamma^{\mu} E_L \right), 
\nn \\ 
 {\mathcal{L}_{\rm{Z}}} &=& 
   \frac{g}{2c_\ww} Z_{\mu} \Big( \bar{N}_L \gamma^{\mu} N_L + ( 2s_\ww^2 -1) \bar{E}_L 
                                   \gamma^{\mu} E_L + 2s_\ww^2\bar{E}_R \gamma^{\mu}E_R \Big) \,, 
\nn \\
{\mathcal{L}_{\rm{A}}}  &=& -eA_{\mu} \bar{E} \gamma^{\mu}E \,,
\label{eq:Lcur}
\end{eqnarray} 
where $g$ is the weak coupling constant and $c_\ww \equiv \cos\theta_W$ and $s_\ww \equiv \sin\theta_W$, where $\theta_W$ is the Weinberg angle. 

The Lagrangian for the left handed SU(2) adjoint triplet
$\omega = (w^1,w^3,w^3)$ and the right handed singlet $\beta^{\dagger}$ Weyl fermions is
\begin{equation}
{\mathcal{L}_{\rm{Ad,g}}} =  i\omega^{\dagger}\bar{\sigma}^\mu D_\mu \omega + i\beta\sigma^\mu\partial_\mu \beta^{\dagger} \,.
\label{eq:Lad}
\end{equation}
Here $\sigma^\mu \equiv (1, \vec{\sigma})$ and $\bar{\sigma}^\mu = \sigma_\mu$, where $\sigma^i$ are the usual Pauli matrices, and the covariant derivative is $D^{ac}_\mu =  \partial_\mu \delta^{ac} + g \epsilon^{abc} A^b_\mu$. We can go to the 4-component notation by defining charged Dirac spinors $w_D^{-}=({w}^-_\alpha \,  (w^+)^{\dagger\dot{\alpha}})^T$ and $w_D^{+}=({w}^+_\alpha \, (w^-)^{\dagger\dot{\alpha}})^T$ and neutral Majorana spinors $w_M=({w}^3_\alpha \;(w^3)^{\dagger \dot{\alpha}})^T$ and $\beta_M = (\beta_\alpha \; \beta^{\dagger\dot{\alpha}})^T$,
where $w^\pm=(w^1\mp iw^2)/\sqrt{2}$ are the two-component charge eigenstates. In the 4-component notation the Lagrangian (\ref{eq:Lad}) becomes:
\begin{eqnarray}
{\mathcal{L}}_{\rm{Ad,g}}
&=& i\bar w_D \deldag w_D 
  + \frac{i}{2}\bar w_M \deldag w_M 
  + \frac{i}{2}\bar \beta_M \deldag\beta_M 
\nonumber \\
&+& g\left(
    W_\mu^+\bar{w}_M\gamma^\mu w_D
   +W_\mu^-\bar{w}_D\gamma^\mu w_M
   -W^3_\mu\bar{w}_D\gamma^\mu w_D \right) \,,
\label{gauge_lagrangian}
\end{eqnarray}
where $w_D\equiv w_D^-$. Note that the neutral field $w_M$ does not couple to neutral gauge boson, and that the singlet field $\beta_M$ has no gauge couplings. This interaction pattern has important consequences for the observability of the DM particle in the model.

%
\subsection{Scalar interactions and masses}
%

The masses and mixings of the new fields introduced above are determined by their couplings with the scalar sector, which we represent by an effective SM-like Higgs doublet $H$. First, the gauge-invariant interactions between $H$ and the $4^{\rm{th}}$ family leptons and the neutral singlet are, up to dimension five operators, given by:
\begin{equation}
{\mathcal{L}}_{\rm{4f,H}} 
  = y_E \bar{L}_L H E_R 
  + y_\beta\bar{L}_L\tilde{H}\beta_{R} 
  + \frac{\LNN}{\Lambda}(\bar{L}_L\tilde{H})(\tilde{H}^TL^c_L)+{\rm{h.c.}} \,,
\label{4fhiggsmass}
\end{equation}
where $\tilde{H}=i\tau^2H^\ast$ and $y_E$, $y_\beta$ and $\LNN$ are dimensionless coupling constants and $\Lambda \gg v$ is yet some unknown scale related to the UV complete theory generating the full flavor structure of the model. After symmetry breaking the first Yukawa term generates a Dirac mass $m_E = y_E v/\sqrt{2}$ for the charged lepton $E$, where $v$ is the vacuum expectation value of the neutral composite Higgs field $h$. Second Yukawa term and the non-renormalizable dimension five operator produce Dirac and Majorana mass terms for the neutrino mass matrix.

Note that all interactions in Eqs.~(\ref{eq:Llep}), (\ref{gauge_lagrangian}) and (\ref{4fhiggsmass}) are invariant under $Z_2$ symmetry transformation, where 
\begin{equation}
E\rightarrow -E, \qquad  N \rightarrow -N, \qquad \beta \rightarrow -\beta \quad {\rm and} \quad w\rightarrow -w \,.
\end{equation}
We take this to be an exact symmetry of the model, which then forbids any couplings between the SM particles and the doublet or triplet fields. This symmetry is crucial for the stability of the dark matter in the model. We next allow all couplings between $H$ and the SU(2) adjoint fields, which are consistent with the $Z_2$-symmetry, again up to dimension five operators:
\begin{equation}
{\mathcal{L}}_{\rm{Ad,H}}
 =  y_w \tilde{H}^T\omega L_L 
  + \frac{\Lwb}{\Lambda} \beta H^\dagger\omega H 
  + \frac{\Lww}{\Lambda} H^\dagger\omega\omega H \, + \, {\rm{h.c.}}  \,,
\label{Adhiggsmass} 
\end{equation}
where $\omega\equiv \omega^a\tau^a$ and $\tau^a=\sigma^a/2$ in terms of the Pauli matrices and the scale $\Lambda$ is the same we introduced in Eq.~(\ref{4fhiggsmass}). Finally, we include a gauge- and $Z_2$ symmetric interaction Lagrangian to provide a mass to the singlet field $\beta$~\footnote{We assume that the bare mass term for $\beta$ is zero even in the absence of a protecting symmetry principle. Whether such symmetry exists is related to a broader issue concerning the underlying flavor physics yielding also the effective couplings between the scalars and fermions. This is an interesting topic which, however, is beyond the scope of the analysis carried out in this paper.}: 
\begin{eqnarray}
{\mathcal{L}}_{\beta S} = y_R S \beta\beta + {\rm{h.c.}} 
\label{singlet} 
\end{eqnarray}
After symmetry breaking the interactions (\ref{4fhiggsmass}) and (\ref{Adhiggsmass}-\ref{singlet}) give rise to a $3\times3$ mass matrix for the neutral Majorana particles $N$, $w_M$ and $\beta_M$ (we drop the index $M$ from $w_M$ and $\beta_M$ here):
\begin{eqnarray}
{\mathcal{L}}_{\rm{mass}} 
  = \frac{1}{2} \left( \begin{array}{ccc} \overline{N}_R, \, \overline{w}_R,  \, \overline{\beta}_R \end{array} \right) 
				\left( \begin{array}{ccc} \MNN & \mNw  & \mNb \\
				                          \mNw & \Mww  & \mwb   \\
				                          \mNb & \mwb  & \Mbb \end{array} \right)
				\left( \begin{array}{ccc} N_L \\ w_L \\ \beta_L  \end{array} \right)  + {\rm{h.c.}}
\label{eq:massmatrix}
\end{eqnarray}
where $\MNN = \LNN v^2/\Lambda$, $\Mww = \Lww v^2/4\Lambda$, $\Mbb = \sqrt{2}y_Rv_s$, $\mNw \equiv  y_w v/2\sqrt{2}$, $\mNb = y_\beta v/\sqrt{2}$ and $\mwb = \Lwb v^2/2\Lambda$, where $v_s$ is the VEV of the singlet field $S$. The lightest mass eigenstate of this mass matrix is stable by the $Z_2$ symmetry, and is thus identified as the DM particle. Finally note that the Majorana mass $\Mww$ is simultaneously the mass of the charged adjoint field $w_D$.

%
\subsection{Rotation to the mass eigenbasis}
\label{sec:massmixing}
%

The symmetric mass matrix appearing in Eq.~(\ref{eq:massmatrix}) can be diagonalized by a unitary transformation $U^{T} M U = m$, such that the mass eigenvalues are $m_i\geq 0$. Using the notation $\Omega_L \equiv (N_L ,\, w^0_L, \, \beta_L )^T$ Eq.~(\ref{eq:massmatrix}) can then be written in the form
\begin{equation}
{\mathcal{L}}_{\rm{mass}} \; = \; \frac{1}{2} \overline{\Omega}_R M \Omega_L  
                             +\frac{1}{2} \overline{\Omega}_L M^{\dagger} \Omega_R                            \; = \; \frac{1}{2} \overline{\chi} m \chi \,,
\label{eq:masseigenvalues}
\end{equation}
where $m$ is the diagonal mass matrix with positive mass eigenvalues. The corresponding mass eigenstates are Majorana fields given by $\chi = \chi_L + \chi_R \equiv U^{\dagger} \Omega_L +  U^{T} \Omega_R $\,. This relation can be immediately inverted to give $\Omega_L = U \chi_L$ and  $\Omega_R = U^* \chi_R$. Using these relations and Eqs.~(\ref{eq:Lcur}) and (\ref{gauge_lagrangian}) we then find the weak currents of the heavy leptons and the SU(2) adjoint fermions in the mass eigenbasis:
\begin{eqnarray}
{\mathcal{L}^{\rm{W}}_{\rm{4f}}} 
& = & \frac{g}{\sqrt{2}} W_{\mu}^{-} 
\sum_i U_{1i} \bar{E}_L \gamma^{\mu} \chi_{iL}  + \rm{h.c.}  \, , 
\label{Wcur} \\ 
{\mathcal{L}^{\rm{Z}}_{\rm{4f}}} 
& = & \frac{g}{2c_\ww} Z_{\mu}
\Big(\sum_i |U_{1i}|^2 \, \bar{\chi}_{iL} \gamma^{\mu} \chi_{iL} 
+ \sum_{i>j} \bar{\chi}_{i}(iV_{ij}+A_{ij}\gamma^5)\gamma^{\mu} \chi_{j} \Big) \,,
\label{Zcur} \\
{\mathcal{L}}^{\rm{W}}_{\rm{Ad}} 
& = & g W_\mu^- \sum_i  \overline{w}_D (V_i+iA_i\gamma^5)\gamma^\mu\chi_{i}  + \rm{h.c.} \,,
\label{eq:weakcur-mbasis}
\end{eqnarray}
\vskip-0.3truecm
\noindent where
\begin{equation}
\label{AB}
V_{ij} = {\Im}(U^*_{1i} U_{1j}),  \quad
A_{ij} \equiv \Re(U^*_{1i} U_{1j}), \quad
V_{i} \equiv  \Re (U_{2i})  \;\; {\rm and} \;\; 
A_{i} \equiv \Im(U_{2i}) 
\end{equation}
and $U_{ij}$ are the elements of the diagonalizing matrix $U$. Similarly, from 
Eqs.~(\ref{4fhiggsmass}), (\ref{Adhiggsmass}) and (\ref{singlet}) we can find the Higgs interactions in
the mass eigenbasis:
\begin{equation}
{\mathcal{L}_{h \chi}} = 
 -\frac{gh}{2m_W}  \sum_{i\le j}  
          \bar{\chi}_{i} (S_{ij}  + P_{ij}  \gamma^5)\chi_{j} 
     -\frac{g^2h^2}{4m_W^2}  \sum_{i} \bar{\chi}_{i} (S^{2}_{ii} + P^{2}_{ii} \gamma^5) \chi_{i}
+ \ldots \,.
\label{eq:Higgscoup-mbasis}
\end{equation} 
Here dots refer to terms which do not affect tree level matrix element calculations and $m_W$ is $W^{\pm}$-boson mass. The various mixing angle and mass dependent coefficients are defined as
\begin{eqnarray}
S_{ij}  &=&  - \mNb A_{ij} + (\delta_{ij}-2) \Mbb D_{ij}+ m_i\delta_{ij}  \,,
\nonumber \\
P_{ij} &=&  - \mNb i V_{ij} - i(\delta_{ij}-2) \Mbb E_{ij}  \,,
\nonumber \\
S^{2}_{ii} &=& - \mNb A_{ii} - \sfrac{1}{2} \Mbb D_{ii} + \sfrac{1}{2} m_i  \,,
\nonumber \\
P^{2}_{ii}&=& - \mNb i V_{ii} +\sfrac{i}{2} \Mbb E_{ii}  \,,
\label{eq_cfactors}
\end{eqnarray}
where $m_i$ is the i'th mass eigenvalue. The projection factors $V_{ij}$ and $A_{ij}$ are as defined in Eq.~(\ref{AB}) and 
\begin{equation}
D_{ij} \equiv \Re(U_{3i} U_{3j}) \quad {\rm and} \quad 
E_{ij} \equiv \Im(U_{3i} U_{3j}) \,.
\label{DE}
\end{equation}
Equations (\ref{Wcur}-\ref{DE}) contain all information needed to calculate the WIMP interaction rates relevant for the relic density and direct detection analyses.

%
\subsection{Renormalizable and anomaly free implementation}
\label{sec:MWTC}
%

The dark matter model discussed above is not a consistent extension of the SM on its own as 
it suffers from quantum anomalies. However, the low energy Lagrangian (\ref{eq:Ltot}) can be embedded {\em e.g.} into the context of a renormalizable, anomaly free TC model where electroweak symmetry is broken dynamically.  One possible realisation is to take this to be the minimal walking TC, and this possibility was explored in ref.~\cite{Kainulainen:2013sva} (see also \cite{Kainulainen:2006wq,Kainulainen:2009rb,Kainulainen:2010pk}). Here, to illustrate different possibilities, we consider an alternative realization. The complete list of new fields and their quantum number assignments is shown in table \ref{chargeassignments}. 
\begin{table}[htb]
\begin{centering}
\begin{tabular}{|l||c|c|c|c|| c|}
\hline
& SU(3)$_c$ & SU(2)$_L$ & U(1)$_Y$ & SU($N_{\textrm{TC}}$) & $Z_2$ \\
\hline
$L_L$ & 1 & 2 & -1/2 & 1 & -1 \\
$E_R^c$ & 1 & 1 & 1 & 1 & -1\\
$\omega$ & 1 & adj. & 0 & 1 & -1 \\
$\beta$ & 1 & 1 & 0 & 1 & -1 \\
$\tilde{g}$ & adj. & 1 & 0 & 1 & -1\\
\hline\hline
$Q_L$ & 1 & 2 & 1/6 & 3 & 1\\
$U_R^c$ & 1 & 1 & -2/3 & 3 & 1\\
$D_R^c$ & 1 & 1 & 1/3 & 3 & 1\\
$\eta_1$ & 1 & 1 & 0 & 3 & -1\\
$\eta_2$ & 1 & 1 & 0 & 3 & -1\\
$\tilde{G}$ & 1 & 1 & 0 & adj. & -1\\
\hline
\end{tabular}
\caption{The table shows the new states added to SM, and their charge assignments under the SM gauge group and the technicolor gauge group which we will consider to be SU(3). Also shown is the discrete matter parity which is even for the SM matter fields.}
\label{chargeassignments}
\end{centering}
\end{table}
In addition to the fields introduced earlier ($L_L$, $E_R^c$, $\omega$ and $\beta$), 
the table shows the new strongly coupled sector responsible for the electroweak symmetry breaking.
Of the new fermion fields, the techniquarks $Q_L$, $U_R^c$, $D_R^c$, $\eta_1$ and $\eta_2$,
are gauged under a new vectorial gauge interaction SU($N_{\rm{TC}}$). We set $N_{\rm{TC}}$=3. 
These elementary fermions form composite fields similar to the mesons and hadrons in QCD. 
We assume that only one doublet of the technifermions is gauged under the electroweak symmetry, while the remaining two Weyl fermions ($\eta_1$ and $\eta_2$) are singlet under all SM charges. We assume that similarly to other SM-singlet fermions also $\eta_1$ and $\eta_2$ are odd under the "matter parity" $Z_2$. Finally, there are two Weyl fermions, $\tilde{g}$ and $\tilde{G}$, transforming in the adjoint representation of SU(3) of QCD and TC, respectively\footnote{For concreteness we have assigned the value $Z_2=-1$ for these fields.}. Both of these fields are assumed to be heavy and decoupled from low energy particle spectrum. These field are relevant neither for the dynamical symmetry breaking nor for the dark matter. However, as we will discuss in the next section, when $\tilde{g}$ and $\tilde{G}$ are included, along with the SU(2)-adjoint field $\omega$, the model also gives rise to excellent gauge coupling unification. 

At high energies the technicolor sector is described by the Lagrangian
\be
{\cal L}_{\rm{TC}}=-\frac{1}{4}F^a_{\mu\nu}F^{a\mu\nu}-\overline{Q}_Li{\Ddag_L} Q_L
-\overline{U}_Ri{\Ddag_R} U_R-\overline{D}_Ri{\Ddag_R} D_R
-\bar\eta\, i\tilde{\Ddag}\,\eta,
\ee
where $F_{\mu\nu}$ is the field strength of the technicolor gauge field and $Q=(U,D)^T$. The covariant derivative $\tilde{D}$ contains only the technicolor gauge field while the covariant derivatives $D_{L,R}$ contain also the electroweak gauge fields. 
At low energies the strong dynamics is described by an effective Lagrangian for composite mesons.
Due to the different $Z_2$ parities of the techniquarks, the low energy composites are 
$\Sigma\sim\bar{Q}Q$ and $\sigma\sim\bar{\eta}\eta$. The former is the effective Higgs doublet which in our case is a composite field, and hence the model does not suffer from hierarchy problem. The field 
$\sigma\sim S+i\pi_s$ is another composite complex scalar, singlet under all SM charges.

The low energy effective Lagrangian is
\be
{\cal L}_{\rm{TC,\,eff}}={\rm{Tr}}D_\mu\Sigma^\dagger D^\mu\Sigma+\partial_\mu \sigma^\dagger \partial^\mu \sigma-V(\Sigma,\sigma),
\ee
where $\Sigma=(\zeta+i\vec \pi \cdot \vec \sigma)/2$ is charged under the electroweak interactions  ($\sigma_i$ are the Pauli matrices) and
\be
V(\Sigma,\sigma)=m^2{\rm{tr}}M^\dagger M+\lambda{\rm{tr}}(M^\dagger M)^2+ \frac{1}{2}\mu_s^2\pi_s^2,
\label{effpot}
\ee
where $M=\Sigma\oplus \sigma$. The real part, $S$, is identified with the field introduced in Eq. (\ref{singlet}) to provide mass to the fermion field $\beta$, while for the pseudoscalar component $\pi_s$ we have included an explicit mass term in  Eq. (\ref{effpot}). Even if the field $\pi_s$ is light with respect to the intrinsic scale $\Lambda_{\rm{TC}}$, it can be heavy with respect to the masses in the dark matter sector. If $\mu_s$ was comparable to, or smaller than the fermionic DM mass, the dark sector of the model would be more complicated, containing both fermionic and bosonic components. While this is an interesting possiblity, we shall here assume that $\pi_s$ is heavy and consequently the DM is purely fermionic. A mass term for $\pi_s$ is expected to arise from the flavor physics providing the masses of SM matter fermions, and their origin can be accounted for by gauge dynamics (as in extended TC models).

As a final remark, we note that with the above particle content the TC sector might be close to conformality \cite{Ryttov:2009yw}. However, coupling between the TC sector and the SM fields via the gauge and Yukawa interactions will move the theory away from the conformal window 
\cite{Kondo:1988qd,Appelquist:1988fm,Fukano:2010yv}. Therefore, we will assume the properties of the TC sector to be QCD-like.

%
\section{Unification}
\label{sec:Unification}
%
\subsection{Unification of the SM coupling constants}

For completeness we will first briefly review the argument for the unification of the SM coupling constants. More details can be found in ref.~\cite{Kainulainen:2010pk}. At one-loop the coupling constant $\alpha_n$ of an SU$(n)$ gauge theory is given by
\begin{equation}
\alpha_{n}^{-1}(\mu) = \alpha_{n}^{-1}(M_Z) - \frac{b_n}{2\pi}\ln
\left(\frac{\mu}{M_Z}\right) \,.
\label{running1}
\end{equation}
The beta function coefficient $b_n$ is:
\begin{equation}
 b_n = \frac{2}{3}T(R) N_{wf} + \frac{1}{3} T(R')N_{cb} -
 \frac{11}{3}C_2(G) \,,
\label{eq:bn}
\end{equation}
where $T(R)$ and and $T(R')$ are the Casimirs of the representation $R$ for $N_{wf}$ Weyl fermions and of the representation $R'$ for $N_{cb}$ complex scalars and $C_2(G)$ is the quadratic Casimir of the adjoint representation of the gauge group. For SM we have three coupling constants corresponding to $n=3,2,1$. Requiring that SM coupling constants unify means that the three couplings are all equal at some scale $M_{\rm U}$: $\alpha_3(M_{\rm U})= \alpha_2(M_{\rm U})=\alpha_1(M_{\rm U})$ with $\alpha_1= \alpha/(c^2\cos^2 \theta_W)$ and $\alpha_2 = \alpha/ \sin^2\theta_W$, where $c$ is a normalization constant that depends on the choice of the unifying group. Here we shall use $c=\sqrt{3/5}$, corresponding to the SM matter unified into SU(5). 

Using Eq.~(\ref{running1}) we can now derive the following relation: 
\begin{equation} 
B \,\equiv\, \frac{b_3-b_2}{b_2-b_1} = 
      \frac{\alpha/\alpha_3 -\sin^2\theta_W}
           {(1+c^2) \sin^2\theta_W-c^2} = 0.721 \pm 0.004\,,
\label{unification}
\end{equation}
where the Weinberg angle $\theta_W$ and weak and strong coupling constants were evaluated at the $Z$-mass scale, using values from ref.~\cite{Agashe:2014kda}: $\sin^2 \theta_W (M_Z) = 0.23126\pm 0.00005$, $\alpha^{-1}(M_Z) = 127.940\pm 0.014$, $\alpha_3(M_Z) =0.1193\pm 0.0016$ and $M_Z = 91.1876\pm 0.0021$ GeV. 

The hypercharge assignment of our model renders the Technicolor sector identical to one extra SM generation from the electroweak interaction viewpoint. In addition we have one strongly interacting adjoint Weyl fermion, which affects the running of the QCD coupling and one weak triplet affecting the running of $\alpha_2$. The group factors are $T(R)=1/2$ for the fundamental representation, $T(G)=n$ for the adjoint SU(N)-representation and $T(R)=c^2Y^2=(3/5)Y^2$ for the $U(1)_Y$ hypercharge gauge group. With $N_g$ generations of ordinary fermions we then find: 
\begin{eqnarray}
b_1 & = & \frac{4}{3}N_g + \frac{4}{3}  \,\,\; = \;b^{\rm SM}_1 + \frac{37}{30}\nonumber\\
b_2 & = & \frac{4}{3}N_g - \frac{14}{3} \,= \;b^{\rm SM}_3 + \frac{5}{2} \nonumber\\
b_3 & = & \frac{4}{3}N_g - 9 \;\;\;= \;b^{\rm SM}_3 + 2 \,.
\label{unif_coeff}
\end{eqnarray}
Note that the differences $b_i - b_j$ are independent of $N_g$, because they can not be affected by states forming complete representations of the unifying gauge group~\cite{Li:2003zh}. It is now clear that the SM does not unify since $B^{\rm SM}_{\rm theory} \simeq 0.53$. However in our model $B^{\rm TC}_{\rm theory} \simeq 0.722$, which is generously within one sigma of the extremely tight constraint (\ref{unification}).  In fact the unification mass and coupling are very precisely determined by the coupling constant unification. If we define a chi-squared function 
\begin{equation}
\chi^2(M_{\rm U},\alpha_{\rm U}) \equiv \sum_{i=1}^3 \frac{( \alpha_i(M_{\rm U})-\alpha_{\rm U})^2}{\Delta \alpha_i^2} \,,
\label{eq:chisquared}
\end{equation}
where $\Delta \alpha_i$ are the observational errors for each coupling given above in the text (these errors propagate essentially as such to the unification scale), we find that at $1\sigma$-level:
\begin{equation}
M_{\rm U} = (2.20 \pm 0.03) \times 10^{15} \,{\rm GeV},  \qquad
{\rm and} \qquad \alpha_{\rm U} = 0.03042 \pm 0.00002 \,.
\label{eq:MUbound}
\end{equation}

In any grand unified theory nucleons are expected to decay via the exchange of gauge bosons with GUT scale masses. Schematically the partial decay width of the proton into a generic channel containing a meson and a lepton is
\be
\Gamma=\gamma_{\rm{QCD}}\gamma_{\rm GUT},
\ee
where $\gamma_{\rm GUT} $ contains the details of the underlying unified theory and $\gamma_{\rm{QCD}}$ contains the QCD parameters and the effective low energy constants parametrizing the hadronic matrix element relevant for the decay in question. For example, for a simple decay mode via a massive gauge boson exchange, assuming $M\approx M_{\rm U}$, $\gamma_{\rm{GUT}}\sim \alpha_{\rm U}^2/M_{\rm U}^4$. The precise details of course depend on the particular GUT model. We do not pursue such model building here; see e.g. \cite{Ross:1985ai,Giudice:2004tc}.  

To obtain a parametric estimate, consider
\be
\tau_N=\frac{1}{\Gamma}\sim \frac{f_\pi^2}{m_N}\frac{M_{\rm U}^4}{\alpha_{\rm U}^2\alpha_N^2},
\ee
where $f_\pi=0.131$ GeV, $m_N$ is the mass of the nucleon and $\alpha_N$ the hadronic low energy constant. This must be determined from the lattice \cite{Aoki:1999tw} and is subject to relatively large uncertainties~\cite{Aoki:2008ku,Aoki:2013yxa}; for $p\rightarrow e^+\pi^0$ the estimates for the value of $\alpha_N$ range from 0.003 GeV$^3$ to 0.03 GeV$^3$ \cite{Aoki:2008ku}. Using the value 0.01 GeV$^3$ compatible with the lattice calculation~\cite{Aoki:2008ku} and numbers from Eq. (\ref{eq:MUbound}) results in $\tau_N\sim 10^{35}$ y, which is compatible with the current bound from the Super-Kamiokande $\tau_N > 10^{34}$ y~\cite{Agashe:2014kda}.

\begin{figure}
\begin{center}
\includegraphics[width=0.47\textwidth]{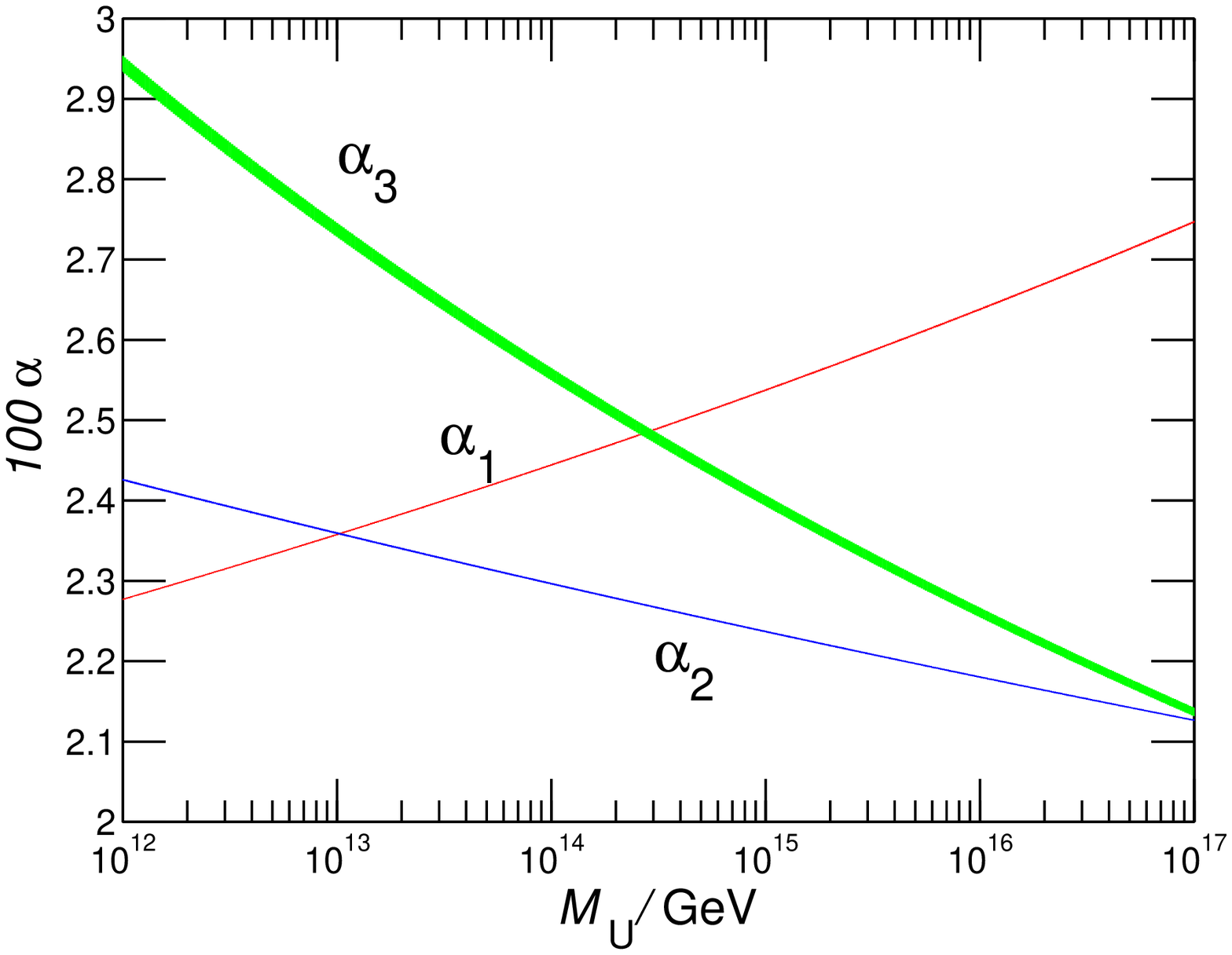}\quad
\includegraphics[width=0.47\textwidth]{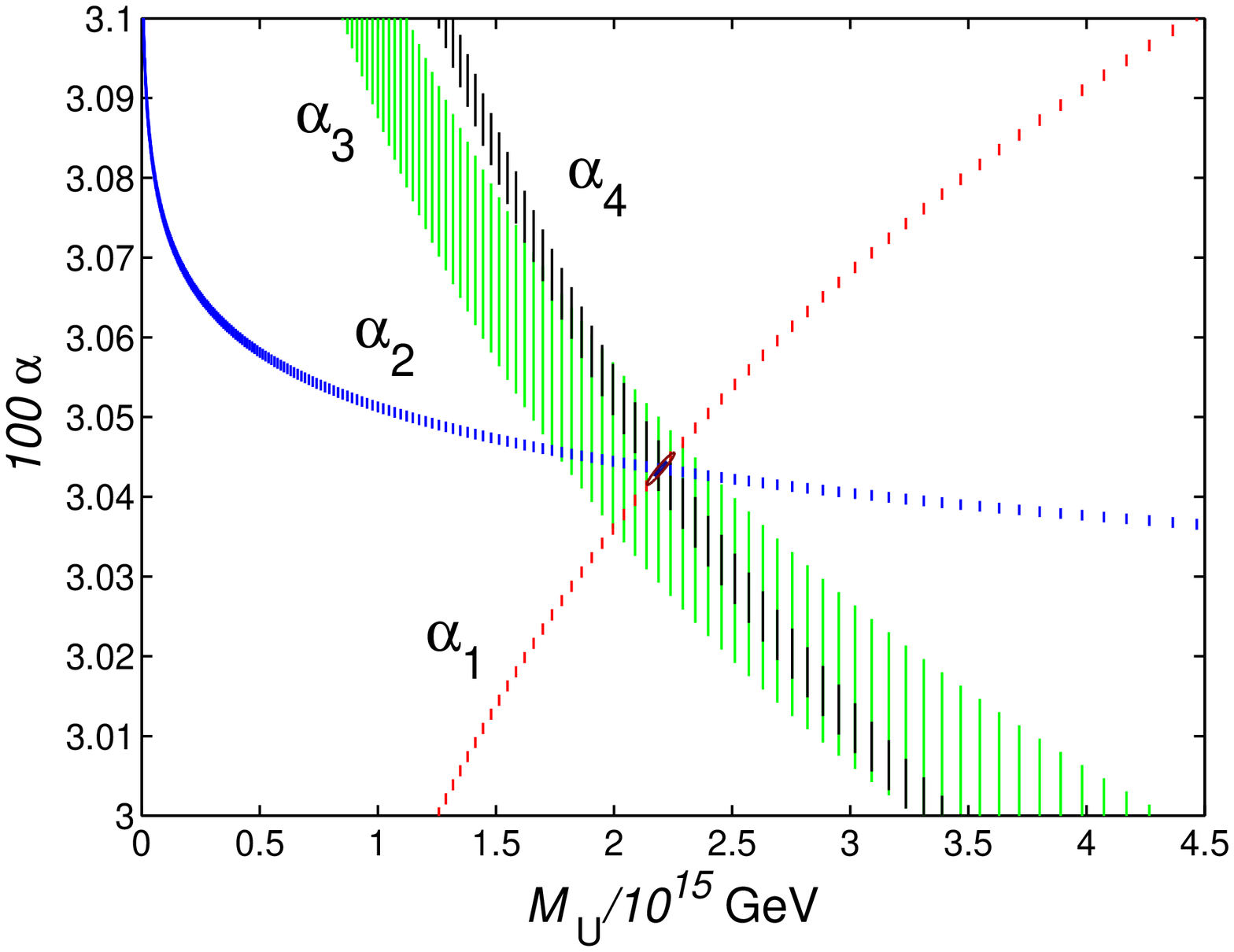}
\caption{Left: the running of the gauge couplings in the SM. Note the logarithmic scale. Left: the running of all four couplings in the MWTC-DM model under consideration, including the TC-coupling $\alpha_4$. The ellipses show the 1- and 2-$\sigma$ contours on unification scale and the unified coupling derived from the $\chi^2$-distribution (\ref{eq:chisquared}). For the TC-coupling $\alpha_4$ we took 331 GeV $< \Lambda_{\rm TC}^{\rm 1-loop}<$ 351 GeV. Note the linear scale on the latter plot.}
\label{fig:unification}
\end{center}
\end{figure}

We plot the running couplings in figure~\ref{fig:unification} for the SM (left panel) and for the current model (right). The latter plot was created with linear $M_{\rm U}$ scale and zoomed to the unification coupling to reveal the almost perfect one-loop unification in our model. The tight error bars on the unification mass and the unified coupling can be used to make a formally accurate prediction for the value of the QCD-coupling at electroweak scale. Running the QCD-coupling backwards from the unification scale gives:
\begin{equation}
\alpha_3(M_Z) = 0.1120 \pm 0.0003 \,,
\end{equation}
which is consistent with but much tighter than the current observational limits\footnote{Note that the running of the SM couplings will be affected by the strongly coupled TC sector at the scales below ${\cal O}({\rm{TeV}})$: the composite spectrum of technihadrons charged under the electroweak interactions will feed into the evolution and may affect the precision of the above result.}.
 
%
\subsection{Unification of all couplings including $\alpha_4 \equiv \alpha_{\rm TC}$}
%

Above we considered only the unification of the SM coupling constants. In our model we have an additional gauge coupling related to the strong Technicolor interactions and it would be more satisfying to have a unification of all four coupling constants. We now show that this indeed quite naturally takes place in our model.\footnote{The results are dependent on the normalization of the hypercharge, i.e. the factor $c$. In principle its value is determined by the particle content and the structure of the unifying algebra; we assume the value $c=\sqrt{3/5}$ throughout here.} 

 With the current particle content, shown in Table~\ref{chargeassignments} we have:
\begin{equation}
b_4 = (4+2) \times\frac{1}{3} + 2 - 11 \; = \; - 7 \,.
\end{equation} 
Let us now require that $\alpha_2$ and $\alpha_4$ unify at $M_{\rm U}$. This implies that:
\begin{equation}
C \,\equiv\,  \frac{\sin^2\theta_W-\alpha/\alpha_4}
           {c^2-(1+c^2) \sin^2\theta_W} = \frac{b_4-b_2}{b_2-b_1} \,.
\label{eq:Cparam}
\end{equation}
Theoretically, the particle content of the model gives:
\begin{equation}
C_{\rm th}  = \frac{b_4-b_2}{b_2-b_1} \approx 1.0556 \,.
\label{eq:Cth}
\end{equation}

The ``experimental" value of $C$ depends on $\Lambda^{\rm 1-loop}_{\rm TC}$, defined as the scale at which the inverse 1-loop coupling vanishes: $\alpha^{-1}_4(\Lambda^{\rm 1-loop}_{\rm TC})\equiv 0$. Using $\Lambda^{\rm 1-loop}_{\rm TC} = M_Z$ gives $C_{\rm exp}^{M_Z} \approx 1.0055 \pm 0.0006$, which is formally about 90 sigma's away from the theoretical value. Since $C_{\rm exp}^{M_Z}<C_{\rm th}$ and $b_4$ is negative, the problem is that $\alpha_4$ {\em undershoots} the unification value. We can improve the situation either by adding new fermion or boson fields to the particle spectrum to achieve a slower running, or by use of a larger $\Lambda^{\rm 1-loop}_{\rm TC}$ to get an effectively positive $\alpha_4^{-1}(M_Z)$ initially in Eq.~(\ref{eq:Cparam}). Let us consider the latter option. From Eqs.(\ref{eq:Cparam}-\ref{eq:Cth}) it is easy to see that the scale $\Lambda^{\rm 1-loop}_{\rm TC}$ needed to achieve unification between $\alpha_4$ and $\alpha_2$ is given by:
\begin{equation}
-\frac{\alpha b_4}{2\pi}\log\frac{\Lambda^{\rm 1-loop}_{\rm TC}}{M_Z}
= \big[ C_{\rm th} - C_{\rm exp}^{M_Z} \Big] ( c^2 - (1+c^2)\sin^2 \theta_W ) \,.
\label{running}
\end{equation}
Using the experimental input values for couplings and for the above computed values for $C_{\rm th}$ and $C_{\rm exp}^{M_Z}$, propagating all errors throughout, we find:
\begin{equation}
\Lambda_{\rm TC}^{\rm 1-loop} = 341 \pm 5 \;\, {\rm GeV} \,.
\label{eq:lambdaTC}
\end{equation}
The right panel of figure~\ref{fig:unification} illustrates how the complete 4-coupling constant unification takes place at the scale and the coupling given by Eq.~(\ref{eq:MUbound}), when $\Lambda_{\rm TC}^{\rm 1-loop}$ is chosen according to Eq.~(\ref{eq:lambdaTC}).
Of course the energy scale where TC dynamics becomes strong is above the scale $\Lambda_{\rm TC}^{\rm 1-loop}$ where the 1-loop coupling diverges. For example, for QCD itself, the simple 1-loop running Eq.~(\ref{running}), yields $\Lambda_{\rm QCD}^{\rm 1-loop} \approx 57$ MeV, while the typical scale for QCD is about an order of magnitude higher: $\Lambda_{\rm QCD} \sim 700$ MeV.  If this holds also for TC, then the unification condition Eq.~(\ref{eq:lambdaTC}) predicts that
\begin{equation}
\Lambda_{\rm TC} \sim 3 \;  {\rm TeV} \,.
\label{eq:lambdaTCTRUE}
\end{equation}
This agrees very well with what one would naively expect: if we write $\Lambda_{\rm TC} \approx 4\pi F_\pi$, where $F_\pi$ is the technipion decay constant, then the unification condition (\ref{eq:lambdaTCTRUE}) sets $F_{\pi} \approx 250$ GeV. 

Even though such scalings are simply naive dimensional analysis, it is encouraging that unification of all couplings is not only possible in the model, but also determines the dynamical symmetry breaking to occur at the TeV scale. Even if our analysis is complete at 1-loop level, the results must be taken with a grain of salt: detailed results may be modified by higher order corrections and by threshold effects. Moreover, we have not addressed the underling nature of flavor dynamic which gives rise to the couplings between the effective scalars and fermion fields, as this dynamics is decoupled from the other gauge interactions at 1-loop level. As an initial exploration of this model, our results nevertheless provide an interesting benchmark scenario for more detailed investigations in the future.

%
\section{Experimental constraints on low energy theory}
\label{sec:Analysis}
%

At low energies our model is parametrized by seven dimensionless couplings $y_E$, $y_\beta$, $y_w$, $y_R$, $\LNN$, $\Lww$, $\Lwb$ and three scales $v$, $v_s$ and $\Lambda$. From these one can easily work out the entries in the mass matrix Eq.~(\ref{eq:massmatrix}) of neutral fields and the mass of the new charged state $E$. From naive dimensional analysis we infer that $y_i < 4\pi$ and $\lambda_{ij} < (4\pi)^2$. Furthermore, assuming $v_s \sim v =$ 246 GeV and $\Lambda\sim {\cal O}({\rm TeV})$, we find it reasonable to adopt the following prior ranges for the Lagrangian masses:
\begin{equation}
| M_{ij} | \leq 3000 \;{\rm{GeV}}; \quad
 | m_{ij}  |  \leq 2000 \;{\rm{GeV}} \quad {\rm and} \quad  200\,{\rm{GeV}} \leq m_E \leq 2000\;{\rm{GeV}} \,.
\label{eq:pmrange}
\end{equation}
We scanned this parameter range using Monte Carlo Markov Chain (MCMC) methods~\cite{Kainulainen:2013sva}. For each set of input parameters we diagonalize the DM mass matrix numerically, find the mass eigenvalues $m_i$ and the diagonalizing matrix $U_{ij}$ and identify the lightest eigenstate as the WIMP. We then check that the WIMP is stable, \ie that it is the lightest of all states transforming nontrivially under the $Z_2$ symmetry. 

The data shown in our result figures is compatible with the experimental and observational constraints
from oblique electroweak precision data, the $Z$-boson and Higgs boson invisible decay width limits, cross section constraints from DM direct detection LUX, XENON100 and PICO experiments as well as DM indirect detection constraints from IceCube, Super-Kamiokande and FERMI-LAT telescopes and from the AMS-02 experiment. For the data passing these tests, the DM relic density is calculated numerically and checked to be consistent with the most recent observations~\cite{Planck:2015xua}.  

We do not require that our model provides the total observed abundance of the DM, inferred from the most recent CMB observations: $\ODM h^2 = 0.1193 \, (\pm 0.0014)$~\cite{Planck:2015xua}. Instead, we impose this as an upper bound and compute how large a fraction of the total DM-density each parameter set is able to produce, defined as $f_{\rm rel} \equiv \Omega_\chi h^2/\ODM h^2$. We accept models also with subleading DM in the interval:
\begin{equation}
0.05 \le f_{\rm rel} \le 1.01 \,.
\end{equation}
This criterion affects the direct and indirect DM search constraints on WIMP-nucleon cross sections, which usually are given assuming that $f_{\rm rel}=1$. However, as long as different DM-components are weakly interacting, they all cluster roughly the same way, and a given subleading DM should make up only a fraction $f_{\rm rel}$ of the DM density in all cosmological substructures. We then constrain such subleading WIMPs using a scaled effective cross section~\cite{Cline:2012hg,Cline:2013bln}:
\begin{equation}
\sigma^{\rm eff}_{\rm SD, SI} \equiv  f_{\rm rel}  \, \sigma_{\rm SD, SI} < \sigma_{\rm bnd}\,,
\end{equation}
where $\sigma_{\rm SD}$ refers to spin-dependent and $\sigma_{\rm SI}$ to spin-independent channel and $\sigma_{\rm bnd}$ is the bound from a given experiment. We imposed direct search bounds from  XENON100~\cite{Aprile:2012nq}, LUX~\cite{Akerib:2013tjd} and PICO~\cite{Amole:2015lsj}, as well as the indirect search bounds from IceCube~\cite{IceCube:2011aj,Aartsen:2012kia} and Super-Kamiokande~\cite{Choi:2015ara,Tanaka:2011uf}. For explicit expressions for cross sections and for more detailed discussion of the implementation of these constraints see~\cite{Kainulainen:2013sva}\footnote{Here we have improved our analysis related to IceCube and Super-Kamiokande limits by taking the DM annihilation branching fractions to different channels into account when imposing the constraints. We use $W^+W^-$ limits for annihilation channels $W^+W^-$,$Z Z$ and $Z h$, and $\tau^+\tau^-$ and $b \bar{b}$ limits as they are, including proper branching fractions in all channels.}. Both SI- and SD-constraints are relevant for our model in different regions of the parameter space. However, the SI-constraint is typically the stronger one. There are particular cases where our DM particle has essentially but a dominantly pseudo-scalar coupling to the Higgs boson. In such case the WIMP-nucleus interaction is momentum transfer dependent and strongly suppressed. These solutions may avoid detection by any of the DM search programs currently under construction.

The new doublet and adjoint SU(2) states in our model, as well as the new states in the TC sector are charged under SU(2) and hence contribute to oblique $S$ and $T$-parameters~\cite{Peskin:1990zt}.
Explicit expressions of these contributions can be found in the appendix of ref.~\cite{Kainulainen:2013sva}. Here we use the experimental constraints~\cite{Agashe:2014kda}:
\begin{equation}
S = 0.00 \pm 0.08,  \quad {\rm and} \quad T = 0.05 \pm 0.07  \,.
\label{expST}
\end{equation}
which include a 90\% correlation between $S$ and $T$ as given by~\cite{Agashe:2014kda}.
There are many other bounds coming from collider experiments. First, there is a direct LEPII-bound on any charged particle $i$ coupling to $Z$-boson: $m_i \ge 104.5$ GeV. LHC mass limits, while not as straightforward to implement, are typically much stronger. In our analysis we have used conservative bounds 
\begin{equation}
m_E, \, m_{\omega_D} > 500 {\rm GeV}.
\label{eq:boundonchargeds}
\end{equation}

The $Z$-boson invisible decay width imposes a constraint on any particle with $m < M_Z/2$. The current bound from LEPII is $\Gamma(Z\rightarrow{\rm{inv.}}) = (2.984 \pm 0.008)\Gamma(Z\rightarrow\bar{\nu}\nu)$~\cite{Beringer:1900zz}. As the best fit value is already 2$\sigma$ below the SM prediction, we allow at most one standard deviation from new physics, which implies a bound
\begin{equation}
\delta_Z \equiv |U_{1i}|^4 \,\Big(1 - \frac{4 m^2_i}{m^2_Z}\Big)^{3/2} < 0.008 \,.
\end{equation}
This rules out any WIMP with $\mDM < m_Z/2$ and a significant $N_L$ component.  Furthermore, if the WIMP is lighter than $m_H/2$, then also Higgs could decay to a pair of WIMPs. The invisible Higgs branching fraction $R_I$ is constrained to be~\cite{Giardino:2012dp,Dobrescu:2012td,Giardino:2013bma}:
\begin{equation}
R_{\rm I} \equiv  \frac{\Gamma_{\rm {H,DM}}}{\Gamma_{\rm {H,DM}} + \Gamma_{\rm {SM,tot}}} \lsim 0.17 \,,
\label{eq:stonginvicible}
\end{equation}
where $\Gamma_{\rm {SM,tot}}$ is the total Higgs decay width in the SM and $\Gamma_{\rm {H,DM}}  =  (G_{F} m_H/2 \sqrt{2} \pi)( \,  |S_{ii}|^2 \beta_i^3 + |P_{ii}|^2 \beta_i)$, where $\beta_i \equiv (1- 4m_i^2/m^2_H)^{1/2}$ and the index $i$ refers to the WIMP as the lightest of the mass eigenstates. The bound (\ref{eq:stonginvicible}) assumes SM-like Higgs-gauge and Higgs-fermion couplings. It would be relaxed to $R_{\rm I} < 0.26$, if one allows Higgs and SM gauge fields to have non-SM-like couplings to photons and gluons~\cite{Giardino:2013bma}. However, in our analysis we always use the stronger constraint (\ref{eq:stonginvicible}).

\begin{figure}
\begin{center}
\includegraphics[width=0.70\textwidth]{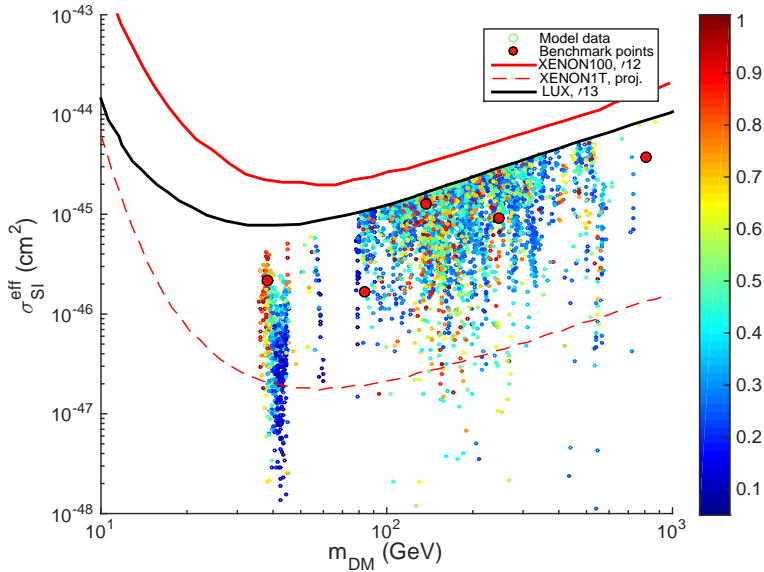}
\caption{Scatter plot of the models passing all existing constraints as a function of the DM-mass and the predicted WIMP-nucleon SI-cross section. Colors represent the value of $f_{\rm rel}$ for each model as indicated by the bar on the right. Also shown are the current XENON100~\cite{Aprile:2012nq}, and LUX~\cite{Akerib:2013tjd} limits as well as the predicted reach of the XENON1T  experiment~\cite{Aprile:2012zx}. 
Large red dots show our five benchmark models.}
\label{Scatter1a}
\end{center}
\end{figure}

Indirect observations are also sensitive on WIMPs annihilating in Galaxy center and in Galaxy halo. However, current limits on signals from neutrino detectors~\cite{Abbasi:2012ws, Lee:2012pz, Aartsen:2013mla} are not stringent enough to be of use here. These annihilations could also create potentially observable flux of gamma-rays in the FERMI-LAT data 
\cite{Buckley:2015doa, Ackermann:2015tah, Ackermann:2012rg, Ackermann:2012qk, Pfrommer:2012mm}. However, our light WIMPs  ($m_{\rm DM} < m_W$) tend to annihilate to lighter fermions in a velocity suppressed $p$-wave via $Z$-boson resonance giving only a very weak gamma-ray signal. Thus, in our model the annihilation cross sections are below the current FERMI-LAT limits~\cite{Buckley:2015doa, Ackermann:2015tah, Ackermann:2012rg, Ackermann:2012qk, Pfrommer:2012mm} in the most constrained low DM mass region. Only the  FERMI-LAT limits from the Milky Way dwarf spheroidal satellite galaxies (dSphs)~\cite{Ackermann:2015zua} are sensitive enough to constrain a small part of our model parameter space. In WIMP mass region $m_W \lesssim m_{\rm DM} \lesssim 100$ GeV these limits \cite{Ackermann:2015zua} cut away a few points.  
For $m_{\rm DM} \gtrsim 100$ GeV these limits have no impact. In all our results we show only data which passes these FERMI-LAT constraints. Finally, even though the annihilation cross section for subdominant DM in dSphs is larger than the canonical thermal relic cross section, the DM density is also expected to be smaller, scaled down by factor $f_{\rm rel}$. This suppresses the flux of gamma-rays by a factor $f^2_{\rm rel}$ making these constraints for subdominant DM even milder than for canonical thermal DM.

Finally, we imposed the constraints for DM annihilation cross section in the $b \bar{b}$ channel from ref.~\cite{Giesen:2015ufa} (left panel Fig.~4), derived from the new AMS-02~\cite{Ams022015} data on the secondary astrophysical antiproton to proton ratio. These limits cut away the few otherwise remaining points in the mass range $m_{\rm DM} \sim 65-80$ GeV.

%
\subsection{Results of a generic MCMC scan}
\label{subsec:scan}
%

In figure~\ref{Scatter1a} we show the distribution of the models that passed all tests in our MCMC runs as a function of the DM-mass, the effective WIMP-nucleon SI-cross section and the relative relic abundance $f_{\rm rel}$, whose value is indicated by the vertical bar to the right of the plot. The advantage of our using the effective cross section here is that one immediately sees how much a given direct search experiment needs to improve its sensitivity in order to rule out a given set of parameters. It is still easy to find acceptable models, in particular with a subleading DM. However, most of the allowed parameter space, including all our benchmark models to be defined below, is within the reach of the next round of the direct search experiments, which improve the current bound on $\sigma_{\rm SI}^{\rm eff}$ by a factor of $\sim 50$.

Let us now very briefly explain the data in figure~\ref{Scatter1a}. When $m_{\rm DM} \lesssim 80$ GeV, the DM particles annihilate into light SM fermions. A vertical cluster of points around $m_{\rm DM} \approx 45$ GeV corresponds to the Z-boson resonance. Another, weaker cluster, around  $m_{\rm DM} \approx 60$ GeV (containing only blue points) corresponds to the Higgs boson resonance. Solutions at range $m_{\rm DM} \sim 65-80$ GeV are exlcuded by AMS-02 constraints as was previously explained. For $m_{\rm DM} \gtrsim 80$ GeV, new annihilation channels, first to $W^-W^+$ and then to $Z Z$, $Z h$, $h h$ and $t \bar{t}$ open up sequentially and begin to dominate the cross section. None of these channels have equally striking signature as do the resonances. 

\begin{figure}
\begin{center}
\includegraphics[width=0.70\textwidth]{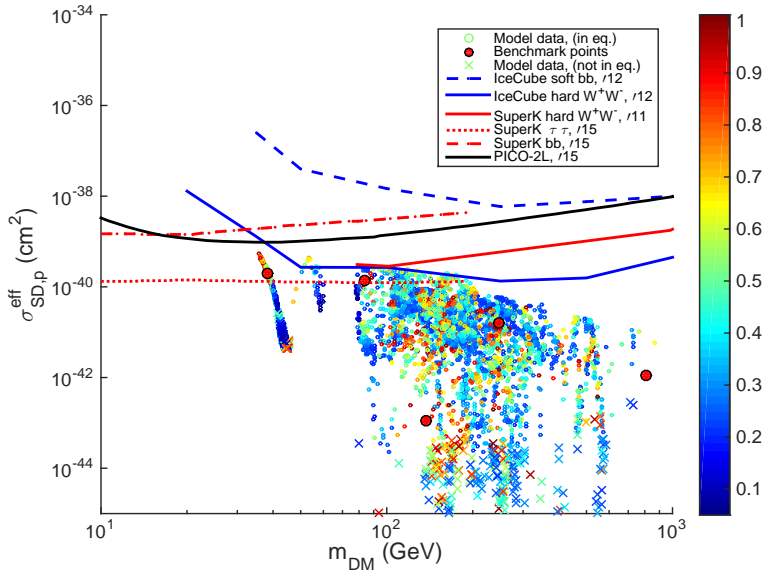} 
\caption{Scatter plot of the models passing all existing constraints as a function of the DM-mass and the WIMP-proton SD-cross section. Shown are also the best current constraints direct~\cite{Amole:2015lsj} (PICO) and indirect searches~\cite{IceCube:2011aj,Aartsen:2012kia} (ICECUBE) and ~\cite{Choi:2015ara,Tanaka:2011uf}(Super-Kamiokande). In models indicated with dots WIMPs would have reached equilibrium between capture and annihilations in the sun and the models indicated with crosses they have not. For more detailed treatment of the indirect observation channels see~\cite{Kainulainen:2013sva}.}
\label{Scatter1b}
\end{center}
\end{figure}
\begin{figure}[t]
\begin{center}
\includegraphics[width=0.70\textwidth]{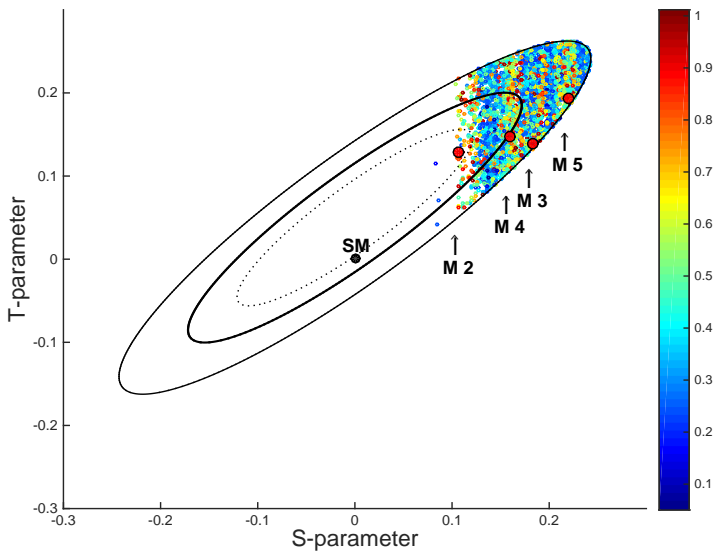}
\caption{Scatter plot of the models passing all existing constraints in the plane of precision electroweak parameters $S$ and $T$. 
The ellipses are the experimental $1 \sigma$, $1.6 \sigma$ and $2.6 \sigma$ confidence contours for $S$ and $T$~\cite{Agashe:2014kda}.
Note that the benchmark model 2 is as good a fit to the data as is the SM.}
\label{Scatter1c}
\end{center}
\end{figure}

In figure~\ref{Scatter1b} we show the projection of accepted points as a function of the spin dependent interaction cross section $\sigma^{\rm eff}_{\rm SD,p}$ and the WIMP mass. 
It is evident that the existing SI-constraints are stronger than the SD-constraints. However, the IceCube and Super-Kamiokande limits, that concern the WIMPs that accumulate in the core of the Sun and then annihilate to $W$-bosons, have some constraining power in the mass region $m_{DM} \lesssim 200$ GeV.  At first sight it seems that the latest Super-Kamiokande results, concerning WIMPs that annihilate into tau leptons, also have some constraining power. Note however, that the points above the Super-Kamiokande $\tau^+\tau^-$-line in Fig.~2 are {\em not} excluded. This is because the Super-Kamiokande constraint assumes that WIMPs annihilate into taus with a branching ratio ${\rm Br}_{\tau\tau} = 1$. 
Here in general, and in the accepted models falling above the Super-Kamiokande constraint in particular, the branching ratio to taus is much less than one. For DM masses below $W^+ W^-$ threshold our WIMPs annihilate dominantly to $b \bar{b}$-channel, and the $\tau^+\tau^-$ branching is only $\sim 0.05$ (for branching ratios in specific benchmark models see table~\ref{tab3}). Furthermore, this ratio only decreases once other annihilation channels open for  $m_{DM} \geq m_W$. 
Due to these small annihilation branching fractions, the Super-Kamiokande limits are currently very little constraining.
\begin{table}[t]
\begin{center}
\begin{tabular}{|c|c|c|c|c|c|c|c|}
\hline
& $M_{\rm NN}$ & $M_{ww}$ & $M_{\beta\beta}$ & $M_{{\rm N}w}$ & $M_{{\rm N}\beta}$ & $M_{w\beta}$  & $m_E$  \\
\hline
{\rm M1}& 3.12-i1.81  & 15.7-i0.37 & 0.60+i0.04  & 0.13+i0.72  & -0.89-i0.11  & -0.45+i0.07 & 7\\
{\rm M2}& 2.18-i0.17  &  8.92+i0.14 &  0.92+i0.04  & -0.02-i0.38 &  -0.29-i0.13  & -0.69+i0.23 &  6.5 \\
{\rm M3}& 7.33-i0.85  & 6.59+i0.49 & 1.42+i0.03  & 0.51+i3.05  & 0.52-i0.41  & -0.70-i0.58 & 12\\
{\rm M4}& 9.39-i0.74  & 7.92-i0.06 & 2.70+i0.14  & -0.21+i2.84  & -0.88-i0.73  & -0.89-i0.53 & 18\\
{\rm M5}& 13.1+i0.14  & 12.7+i0.25 & 8.12-i0.06  & -0.16-i0.13  & -0.38-i0.11  & 0.38+i0.03 & 15\\
\hline
\end{tabular}
\end{center}
\caption{\label{tab2} Shown are the input mass parameters of the benchmark models M1-M5 along with the mass of the new charged doublet state $m_E$. All masses are given in units 100 GeV. Note that the mass of the new adjoint state $\omega_D$ equals with the Majorana mass $m_{w_D}=M_{ww}$.}
\label{tab:smalletaDmodelsA}
\end{table}

%
\subsection{Benchmark models}
\label{subsec:benchmark}
%

\begin{figure}[t]
\begin{center}
\includegraphics[width=0.48\textwidth]{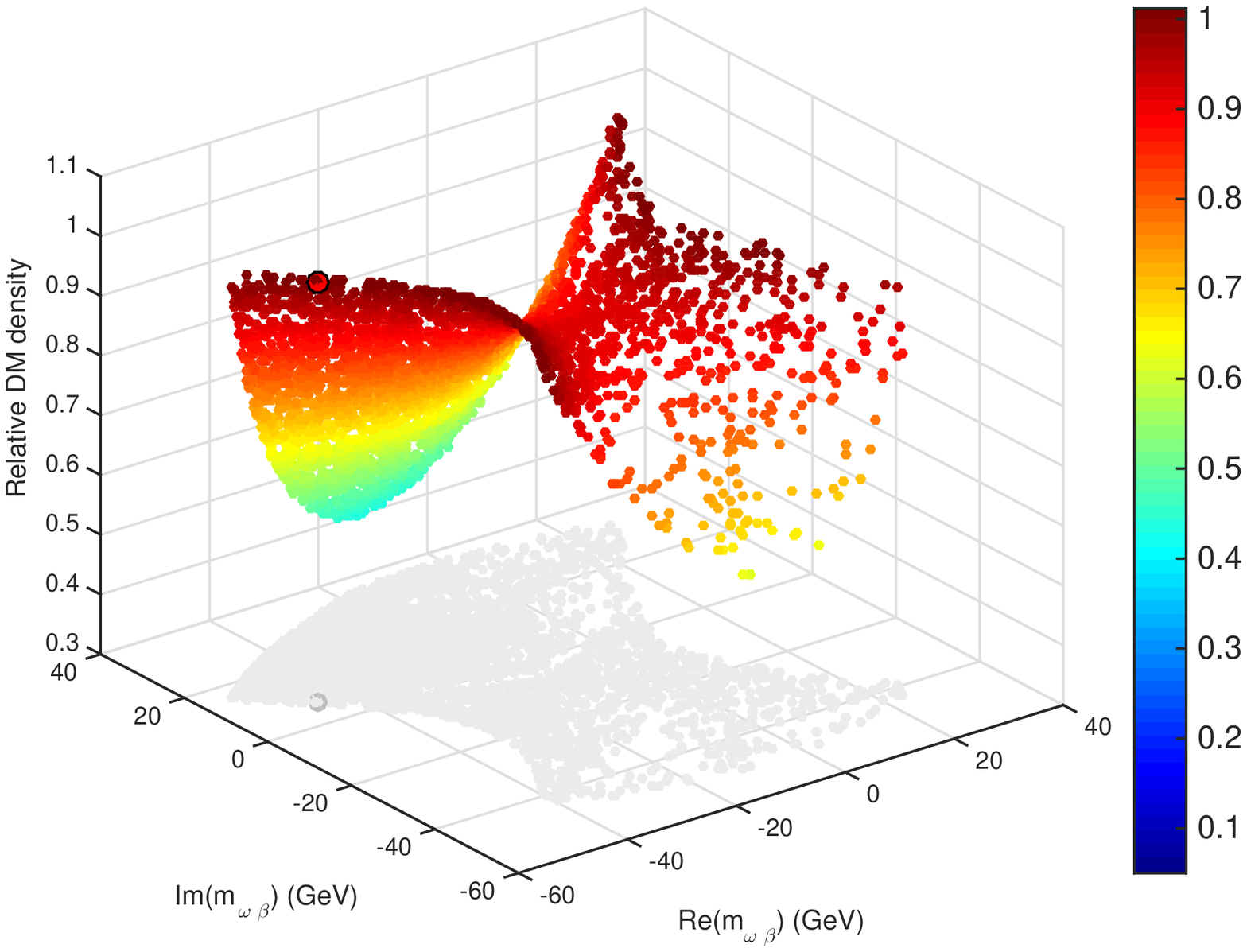} \quad
\includegraphics[width=0.48\textwidth]{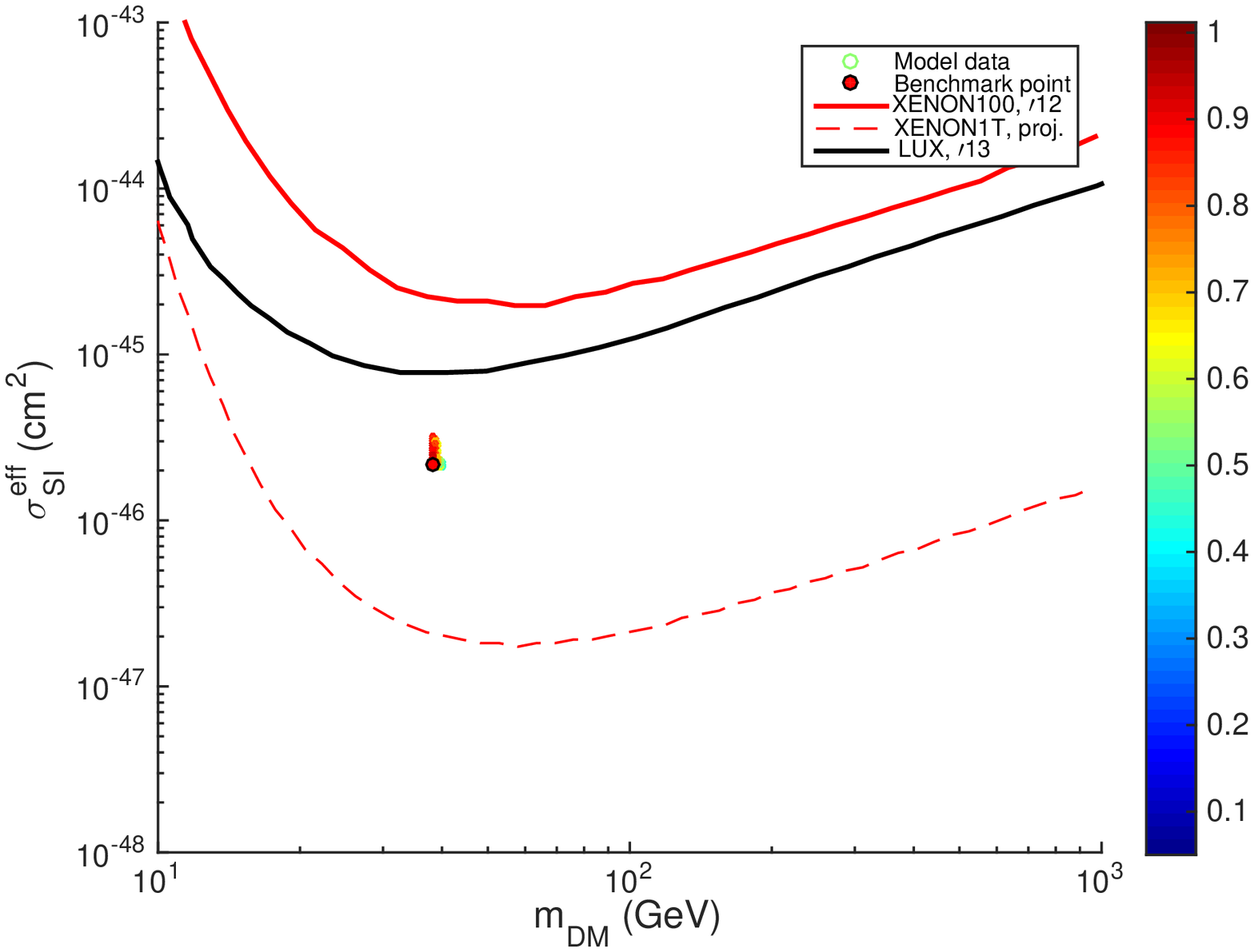} 
\caption{Left: shown is the result of a restricted MCMC-scan in the complex mass parameter $m_{\omega\beta}$ around the benchmark point M1 in Table~\ref{tab:smalletaDmodelsA}. Light gray dots show the projections of the points to the complex mass plane. Right: the projection of these points in $(m_{\rm DM},\sigma_{\rm SI}^{\rm eff})$-plane. } 
\label{Scatter2}
\end{center}
\end{figure}

\begin{figure}[t]
\begin{center}
\includegraphics[width=0.48\textwidth]{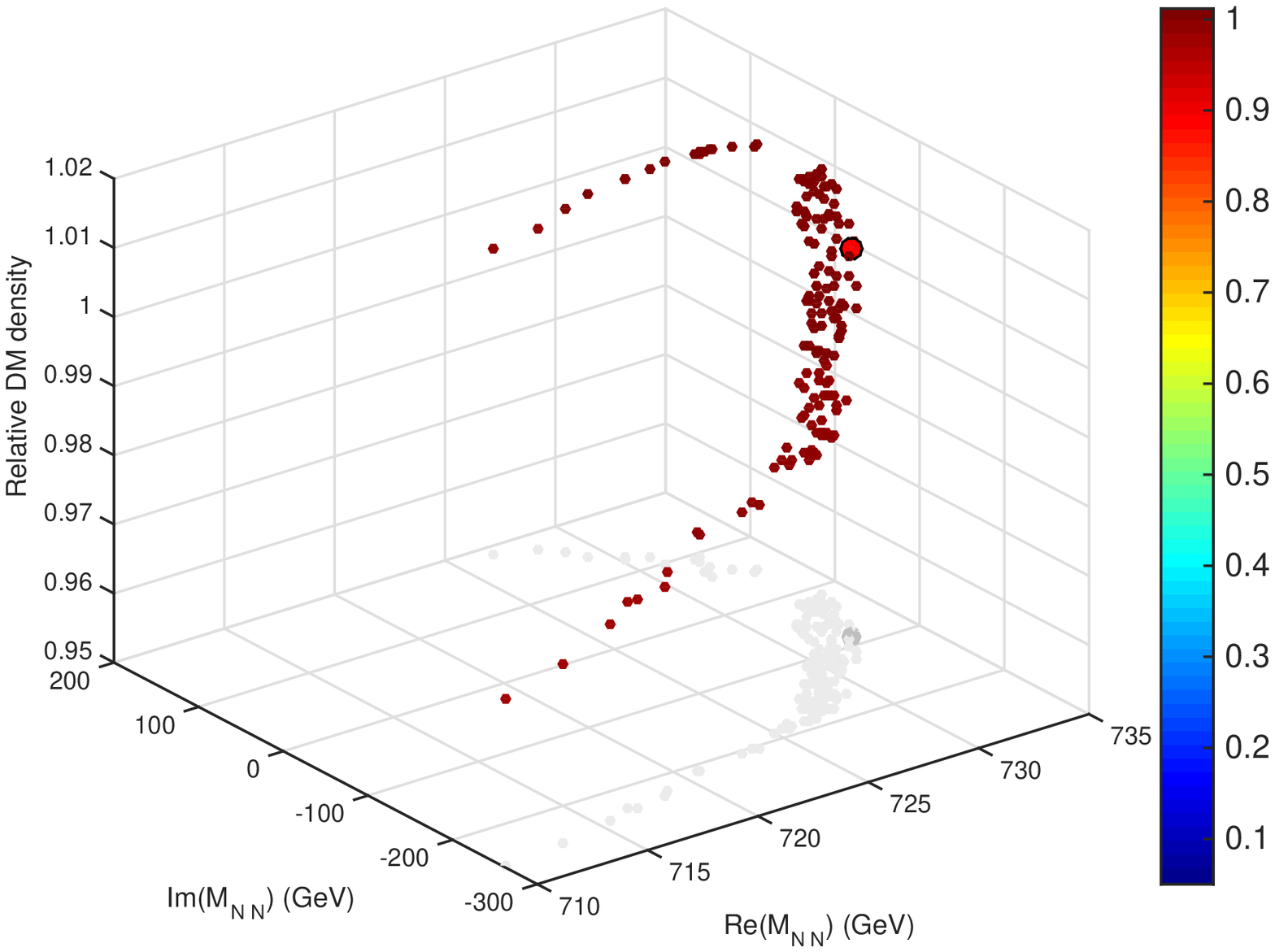} \quad
\includegraphics[width=0.48\textwidth]{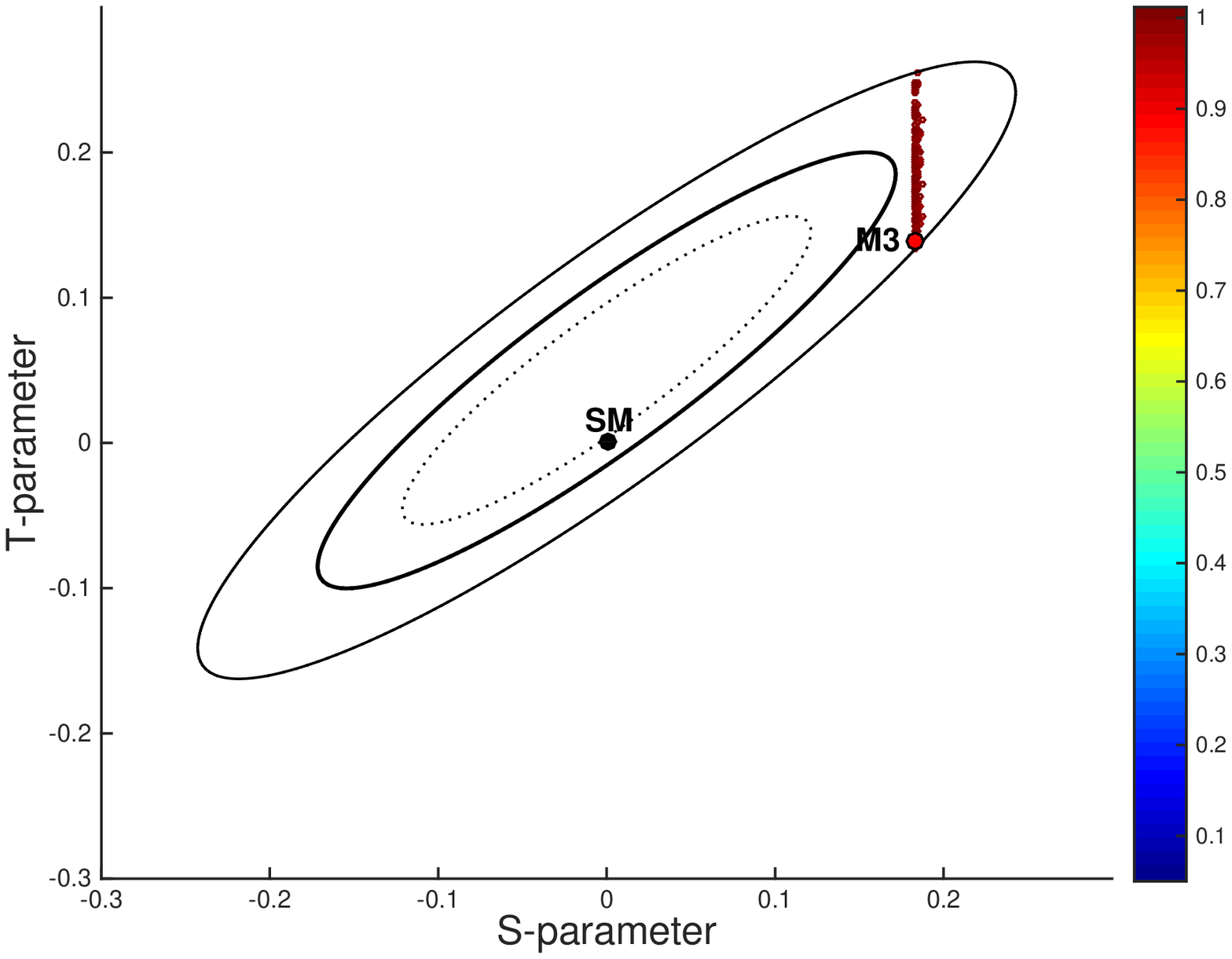} 
\caption{Left: shown is the result of a restricted MCMC-scan in the complex mass parameter $m_{\rm NN}$ around the benchmark point M3 in Table~\ref{tab:smalletaDmodelsA}. Right: the projection of these points in $(S,T)$-plane.}
\label{Scatter3}
\end{center}
\end{figure}

A generic MCMC scan over the entire prior range gives a good idea of the constraining power of the different observations. However, because of the high-dimensionality of the parameter space, such scans do not reveal the finer details of how acceptable models are distributed. In particular it appears that there are but a few sets of parameters that give $f_{\rm rel}\approx 1$. For this reason we selected five benchmark points from the accepted MCMC data sets and made new runs with restricted priors in their neigborhoods. The selected models are labelled as M1-M5 and shown by large red dots in the scatter plots~\ref{Scatter1a}-\ref{Scatter3} and \ref{M_TC}. The corresponding central parameter values are given in table~\ref{tab2}.
In table~\ref{tab:smalletaDmodelsB} we show the ensuing DM mass, relative DM-abundance, precision electroweak parameters $S$ and $T$, the effective WIMP-nucleon cross section $\sigma_{\rm SI}^{\rm eff}$, the tau-branching ratio ${\rm Br}_{\tau\tau}$, the contribution to $Z$-width and the invisible Higgs decay fraction $R_I$ for these models. Note that models 2-5 have large and positive $S$ and $T$-parameters. This is a generic feature in our model, due to the fact that the precision variables, and $S$ in particular, get a large positive contribution from technicolor fields (see table~\ref{chargeassignments}). As is evident from figure~\ref{Scatter1c}, a strict bound on the $S$-parameter $S<0.1$ could rule the model out completely. However, the contributions from flavor extensions are subtle and may quantitatively affect the analysis \cite{DiChiara:2014gsa,DiChiara:2014uwa}. We give no values for $S$ and $T$ for model 1, because our precision data analysis~\cite{Kainulainen:2013sva} is not applicable for $m_{\rm DM}< M_Z/2$. However, we expect that the strongest bound comes from the $Z$-decay width in this region. 

\vskip0.3truecm

\begin{table}[t]
\begin{center}
\begin{tabular}{|c|c|c|c|c|c|c|c|c|}
\hline
 & $m_{\rm DM}$ & $f_{\rm rel}$ & $S$ & $T$ & $\sigma_{\rm SI}^{\rm eff}$ & ${\rm Br}_{\tau\tau}$& $\delta_Z$ & $R_I$  \\
\hline
{\rm M1}& 38   & 1.00   & -     & -      & $2.2\times 10^{-46}$   & 0.051 & 0.0008 & 0.12 \\
{\rm M2}& 84   & 0.96  & 0.11& 0.13 & $1.7\times 10^{-46}$   & 0.0012 & - &-\\
{\rm M3}& 137 & 1.01  & 0.18 & 0.14 & $1.3\times 10^{-45}$  & $7.7 \times 10^{-5}$& - & -\\
{\rm M4}& 246 & 1.01  & 0.16& 0.15 & $9.1\times 10^{-46}$   & $1.3 \times 10^{-5}$& - &-\\
{\rm M5}& 806 & 0.87  & 0.22& 0.19 & $3.7\times 10^{-45}$   & $9.9 \times 10^{-7}$& - & -\\
\hline

\end{tabular}
\end{center}
\caption{\label{tab3} Shown are the values of DM mass $m_{\rm DM}$, relative relic density $f_{\rm rel}$, precision $S$ and $T$ parameters, the effective cross section $\sigma_{\rm SI}^{\rm eff}$, the contribution to $Z$-width $\delta_Z$ and the invisible Higgs decay fraction $R_I$ for the benchmark models. A dash indicates that the bound is not relevant for the model in question.}
\label{tab:smalletaDmodelsB}
\end{table}  

In the left panel of figure~\ref{Scatter2} we show a scan of parameters around the benchmark Model 1. We fixed all parameters as given in table~\ref{tab:smalletaDmodelsA} except $M_{w\beta}$, which was allowed to vary freely in an MCMC scan starting from the benchmark value. This scan reveals a continuous, but constrained domain of parameters giving a right, or very closely right DM abundance. In the right panel we show how these points are all tightly concentrated in the the DM mass-effective SI-cross section diagram. This example shows that the apparent deficit of points with $f_{\rm rel} \approx 1$ in Figs.~\ref{Scatter1a}-\ref{Scatter1c} may give a too pessimistic idea of the density of acceptable models. Figure~\ref{Scatter3} shows a result of an MCMC-run around the benchmark model 3, fixing all parameters but the complex mass $M_{\rm NN}$.  First, we observe that $f_{\rm rel}$ is almost independent of $M_{\rm NN}$. This is because DM is always mostly $\beta$-like and so its mass and couplings do not depend much on the NN-entry of the mass matrix. Second, only a thin line of acceptable solutions are found. The reason for this is the $T$-parameter, which gets a large contribution from the doublet-like states, and the contribution from the $N$-like neutral state must accurately cancel the contribution from the charged $E$-state~\cite{Kainulainen:2006wq}. For a fixed $m_E$ this works only for a very narrow range in the mass of the $N$-like state, which is essentially set by $M_{\rm NN}$. This explains {\rm why} points with $f_{\rm rel} \approx 1$ are relatively sparsely distributed in the generic MCMC plots: our full parameter space has many dimensions (thirteen) and good models are forced to lie on narrow low-dimensional strips, which are hard to locate in a full parameter space scan. However, when good solutions are found, one in general finds continuous sheets of acceptable solutions in their immediate neigborhood. 
\vskip0.3truecm
%

%
\section{Additional bounds: Higgs to $\gamma \gamma$ and a light composite scalar}
\label{sec:Hgg}
%

As already emphasized, our generic DM setting provides a very attractive solution for the particle dark matter problem, the hierarchy problem and the gauge unification. However, before concluding, we still need to discuss two additional constraints. First, the bound on higgs decay to two photons and second, the lightness of the higgs mass in the effective field theory picture underlying the model. We find that both issues can be addressed within the strongly interacting sector of the model and do not have consequences for the DM physics. Even though these bounds are of no concern for our main results on dark matter, we discuss them for the benefit of the overall consistency of the model. A reader not interested in these details may skip directly to the conclusions.

%
\subsection{Higgs to $\gamma\gamma$}
%

There are a number of new particles in our model, which can influence the Higgs boson decay widths. As we have already discussed, if our DM is light enough, $m_{\rm DM}< m_h/2$, it opens up a new invisible Higgs decay channel, which would change the predicted higgs branching ratios. In addition, all new charged particles contribute to Higgs decays to two photons via loop corrections.

The contribution from new TC singlets, the new heavy 
electron $E^\pm$ and the new charged $w_D^{\pm}$ state nested in the SU(2) triplet, are easy to compute. The sector which is not singlet under TC is more model dependent. 
To estimate this effect we consider the contributions of heavy massive resonances within the sigma model-like effective theory describing the Higgs sector at low energies.
The loop corrections induced by vectors to $\gamma\gamma$-amplitude are much larger than those from scalars with a similar coupling strength, and so vector resonances may easily dominate the effective $H\gamma\gamma$ coupling. Concretely, we consider a simple setup with one extra charged vector resonance, described as a new effective massive $W'$-boson. The relevant effective Lagrangian then is 
\begin{eqnarray}
{\mathcal{L}}_{\rm{eff,H}} &=&\frac{2 m^2_W c_W}{v}hW^-_{\mu}W^{+\mu} + \frac{m^2_Z c_Z}{v}hZ_{\mu}Z^{\mu} 
\nonumber \\
&& -\sum_f  \frac{m_f c_f}{v}h\bar{f}f  +  \frac{2 m^2_{W'} c_{W'}}{v}hW'^-_{\mu}W'^{+\mu}
 \,.
\end{eqnarray}

The sum $f$ runs over the charged SM fermions and the new charged techni-singlet fermions. 
In practice these are the top quark, the new heavy electron $E$ and the $w^{\pm}$ states. The effective fermion and EW vector boson couplings of the Higgs are denoted by $c_f$ and $c_W$ so that for SM we have $c_f=c_W=1$.  In  our model $c_E =1$ and $c_{w^{\pm}}=2$, 
while the couplings to the top quark and $W$-bosons must be inferred from the LHC data. Generally, on the basis of extrapolating from QCD and explicit model calculations \cite{Belyaev:2013ida, DiChiara:2014uwa}, it is expected that in this type of a model $c_W = c_Z \approx 1$ and $c_{t} \approx 1$ which is also confirmed by fits to the LHC data ~\cite{Belyaev:2013ida}.\footnote{Note that the Higgs coupling to gluons is SM-like, since we assume that all the new particles coupling with the Higgs are color singlets.}

For simplicity and for less model dependence we have written the Lagrangian in terms of mass eigenstates. We also neglected the mixing of $W$-and $W'$-bosons, which in general could lead to a correlation between $W$-and $W'$-boson couplings. With these assumptions all non-conventional effects due to charged resonances are modelled by the coupling factor $c_{W'}$.

The full 1-loop decay width of Higgs to two photons is
\begin{eqnarray}
\label{GHgg}
\Gamma_{\gamma \gamma} = \frac{m^3_H}{4 \pi v} (g_{H\gamma \gamma })^2,
\end{eqnarray}
where the effective coupling is given by
\begin{eqnarray}
g_{H\gamma \gamma } = \frac{\alpha}{8 \pi} \left| \sum_f c_f N^f_c  Q^2_f F_{1/2}(\tau_f)+ c_WF_1(\tau_W) + c_{W'}F_1(\tau_{W'})\right|.
\label{gSMHgg3}
\end{eqnarray}
The color factors of non-standard model fields are $N^E_c = N^{w^{\pm}}_c = 1$ and the loop factors $F_i(\tau_j)$ for fermions ($i=1/2$) and for vector bosons ($i=1$) are standard,
\begin{eqnarray}
F_{1/2}(\tau_{f}) &\equiv& -2\tau_f \left[1+ (1-\tau_f)f(\tau_f) \right] , \nonumber \\
F_{1}(\tau_W) &\equiv& 2+3\tau_W +3\tau_W(2-\tau_W)f(\tau_W),
\end{eqnarray}
where
\begin{equation}
f(\tau_{j}) \equiv \left\{ \begin{array}{ll}
\arcsin^2 \frac{1}{\sqrt{\tau_j}} & {\rm if} \,\, \tau_j \geq 1 \\
 -\frac{1}{4}\left[ \log \frac{1+\sqrt{1-\tau_j}}{1-\sqrt{1-\tau_j}} -i \pi \right]^2  & {\rm if} \,\, \tau_j < 1 \end{array} \right.
\end{equation}
and $\tau_j \equiv 4 m^2_j/m^2_H$. For $\tau_j$ larger than unity, the loop-factors quickly reach the asymptotic values $F_{1/2}(\infty) = -4/3$ and $F_{1}(\infty) = 7$. Using the asymptotic value is a good approximation for all new fermions and even for the top quark the relative error of this approximation is about three per cent. Consequently, since the techni-resonances are expected to to be in ${\cal O}$(TeV) mass range, the exact value of the $W'$ mass is not relevant and $F_{1}(m_{W'})\approx 7$.

\begin{figure}
\begin{center}
\includegraphics[width=0.60\textwidth]{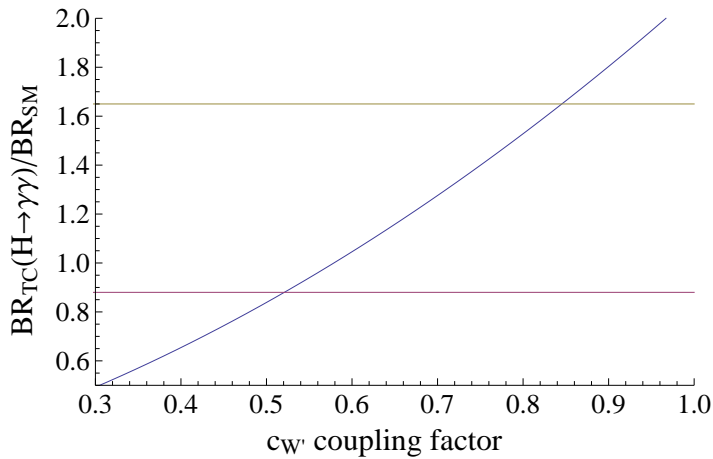} 
\caption{Shown is the $2\sigma$-limit for the coupling factor $c_{W'}$. The allowed region is between the horizontal lines. The factor $c_{W'}$ is the effective coupling between techni-vector resonance and the techni-Higgs.
}
\label{Hgamma}
\end{center}
\end{figure}

The effective coupling $c_{W'}$, can now be constrained  by existing bounds on the ratio of the Higgs branching fractions to two photons~\footnote{Recent results from Atlas collaboration~\cite{Aad:2015lha} favor a slightly larger $\gamma\gamma$-branching  fraction, ${\rm BR}_{\gamma\gamma}/{\rm BR}^{\rm SM}_{\gamma\gamma} > 1$. As is evident from Fig.~\ref{Hgamma}, this situation is naturally realized in our model.}, \ie ${\rm BR}_{\gamma\gamma}/{\rm BR}^{\rm SM}_{\gamma\gamma}$. We extracted a $2\sigma$-limit for this quantity from the left panel of Fig.~5 of ref.~\cite{Giardino:2013bma}. The experimental bound is then converted to a constraint on $c_{W'}$ assuming no invisible decay channels. The result is shown in Fig.~\ref{Hgamma}. The rather loose bound $0.52 \lsim c_{W'} \lsim  0.85$ indicates that our model can 
 be made consistent with the data by rather modest and reasonable assumptions about the structure of the effective low-energy theory.
 
Another possibility would be to include the TC dynamics by considering the elementary fields, i.e. charged techniquarks, inside the loop and interacting with the TC-Higgs. 
We have checked, adapting the resuls of~\cite{Belyaev:2013ida} to our SM-like hypercharge convention, that that this leads to quantitatively similar and consistent results with the effective model approach we have detailed here.

%

%
\subsection{Dynamical Higgs boson mass scale}
\label{sec:dynamic}
%

Then we discuss how a light composite Higgs relates to the intrinsic dynamical scale of the model. 
The traditional expectation of a heavy scalar resonance assumes TC dynamics in isolation. 
It is known that couplings with the EW gauge currents and extended flavour sectors, in particular the top quark, affect this conclusion~\cite{Foadi:2012bb,DiChiara:2014gsa,DiChiara:2014uwa}. In addition, our model contains several new heavy fields that couple to the Higgs and contribute to its vacuum polarization. We estimate this effect quantitatively by use of a perturbative one-loop approximation. As usual, we induce a cut-off $\Lambda$ to evaluate the relevant 1-loop integrals. However, since $\Lambda$ is not very large here, we keep nonzero masses for the heaviest particles in the loops. This procedure gives:
\begin{eqnarray}
m^2_{H}  &=& (m^{\rm TC}_H)^2 
+ \left[ 6 m^2_W + 3 m^2_Z - 12  m^2_t  \right] \frac{\Lambda^2}{16 \pi^2 v^2}  
\nonumber \\
                  &-& 4 \sum_i f_i m^2_i  \frac{\Lambda^2}{16 \pi^2 v^2} 
                  \left( 1 - \frac{m^2_i}{\Lambda^2} \log \frac{\Lambda^2}{m^2_i}\right).
\label{mHTC}
\end{eqnarray}
Here $m^{\rm TC}_H$ is the intrinsic dynamical TC Higgs boson mass, which we attempt to estimate. The cut-off dependent part in the first line includes corrections from the relevant SM particles and the sum in the second line the corrections from the new heavy fermions.  These include the charged fermions $E$ and $\omega^{\pm}$ with factors $f_E =1$ and $f_{\omega^{\pm}} = 4$ and the two heaviest neutral Majorana fermions with a factor $f_{\chi_i}=4$ for each\footnote{The factor $f_i = 4$ follows from the expansion of the Higgs interaction term $(1+ h/v)^2 \rightarrow 2 h/v$. This factor of two in the coupling gives the factor of four in the loop.}. We approximate the masses of the neutral states by $\MNN$ and $\Mww$, which is a reasonable approximation, as the Dirac mass terms are typically small compared to the diagonal Majorana masses in the original WIMP mass matrix. Thus the mixings are small and finally the contribution from the relatively light WIMP is suppressed compared to the other two Majorana states. 
\begin{figure}
\begin{center}
\includegraphics[width=0.7\textwidth]{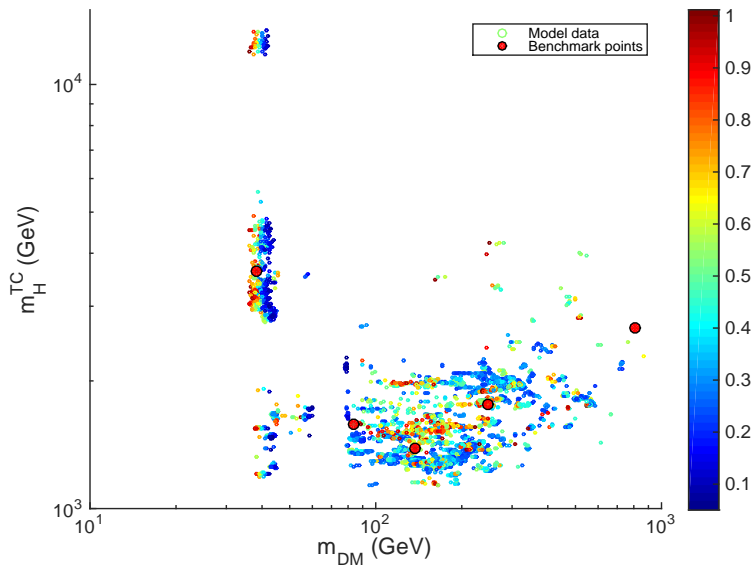}  
\caption{Scatter of the one-loop estimate for the intrinsic TC Higgs mass $m^{\rm TC}_H$ for our accepted models.  The colour coding again indicates the relative DM density.
}
\label{M_TC}
\end{center}
\end{figure}

The cut-off is provided by the TC-scale, which here, consistently with the unification: $\Lambda \sim \Lambda_{\rm TC} \approx 3$ TeV (see Sec.~\ref{sec:Unification}). For concreteness we used $\Lambda = 1.5$ TeV. We can now turn Eq.~(\ref{mHTC}) around, setting $m_H = 125$ GeV and solve it for the dynamical Higgs mass $m^{\rm TC}_H$ for all models in our data set. The result is shown in Fig.~\ref{M_TC}. Models with $2m_{\rm DM} > M_Z$ give $m^{\rm TC}_H \sim $ 1-2 TeV. This is of the right order of magnitude, since one would expect that in a conventional technicolor model $m^{\rm TC}_H < 4 \pi v \approx$ few TeV. For models with $m_{\rm DM}<M_Z/2$ the small mixing approximation made in using Eq.~(\ref{mHTC}) breaks down and the results for $m^{\rm TC}_H$ cannot be trusted. Anyway, these arguments strongly suggest that the Higgs boson can indeed be naturally light in our model.

%
\section{Conclusions}
\label{sec:conclusions}
%

We have analysed a model for dynamical symmetry breaking, dark matter and gauge coupling unification in light of the most recent observational and experimental data. The model setting is nontrivial, but very tightly constrained theoretically. Since the model does not contain any light fundamental scalars, there is no hierarchy problem. We have shown how the model gives rise to perfect 1-loop unification of all gauge couplings, including the new technicolor interaction, at a common unification scale $M_{\rm U} = 2.2 \times 10^{15}$ GeV and the unified coupling $\alpha_{\rm U} \approx 0.0304$. Moreover, unification determines the scale of the 1-loop IR-pole of the TC-coupling $\Lambda_{\rm TC}^{\rm 1-loop} \approx 340$ GeV. By the QCD analogue, this is consistent with the naive expectation of the the TC interactions becoming strong around the scale $\Lambda_{\rm TC} \sim {\cal O}({\rm TeV})$. Thus unification is not only possible, but it actually supports the existence of a TeV-scale strongly interacting dynamical sector.

The essential part of the spectrum are the fermion fields transforming under adjoint representations of the gauge group. In terms of their quantum numbers these fermions are identical to the gauginos which arise in supersymmetric setting. Hence, it is natural to entertain the thought that the model  is a low energy realization of a supersymmetric theory. The existence of a strongly coupled sector responsible for the electroweak symmetry breaking naturally decouples supersymmetry breaking from the electroweak physics, hence removing the little hierarchy problem \cite{Dine:1981za,Antola:2011bx,Antola:2013fka}. Consequently the scalar superpartners can all be very heavy with masses around or above the unification scale $M_{\rm U}$. More detailed model building and investigation of resulting phenomenology provide interesting further research prospects. It would be also desirable to have more detailed theory which explains the effective scalar-fermion couplings in our low energy lagrangian and the emergence of fermion mass patterns. The details of underlying flavor physics likely require an extensions of the technicolor gauge dynamics. 

In this paper our main analysis concerned the dark matter sector of the model.
The dark matter candidate arises from mixing of three neutral fields: one gauge singlet, 
a neutral members of an SU(2)-doublet and an SU(2) triplet. The stability of the lightest of these fields is guaranteed  by a discrete $Z_2$-symmetry. The most essential parameters for the dark matter in model are the entries in the effective mixing mass matrix of these neutral states.
We performed a generic MCMC scan of the model parameter space constraining the model by the most recent bounds following from the accelerators as well as direct and indirect DM-searches. 

We also introduced several explicit benchmark cases to illustrate typical features of viable models and performed limited range MCMC runs in the neighborhood of the benchmark points to study the allowed phase space in more detail. We found that there are large continuous regions of parameters for which the model can provide a naturally stable DM particle with a mass in the range $m_{\rm DM} \sim 30 - 800$ GeV. 

We conclude that the model is viable in light of existing data from collider experiments and cosmological and astrophysical observations. However, future experiments have excellent possibilities to probe the model further: Most of the available parameter space is within the reach of the next generation of DM search experiments. Also a significant shift of the observed precision electroweak parameters towards their SM-values, in particular of the Peskin-Takeuchi $S$-parameter, could rule out the model as a source for DM. Finally, the experimental results on the proton decay will constrain the unification aspects of the model. Of course the interesting possibility is that one or the other of these observations would provide the first evidence of particles compatible with the low-energy spectrum predicted by the model.

%
\section*{Acknowledgements}
\label{sec:acknowledgements}
%

We acknowledge the financial support from the
Academy of Finland, projects 278722 and 267842.

%
\bibliography{JCAP_KTV5.bib}
%

\end{document}